\documentclass[acmtog, authorversion]{acmart}

\AtBeginDocument{%
  \providecommand\BibTeX{{%
    \normalfont B\kern-0.5em{\scshape i\kern-0.25em b}\kern-0.8em\TeX}}}
    
\usepackage{caption}
\usepackage{subcaption}

\usepackage{longtable}
\begin{document}

% \title[Exploring the Design Space of Personalized ST]{A Metadata Analysis \& Exploration of the Design Space of Personalized ST for Health Behavior Change}
\title[12 Years of Self-tracking Research from a User Diversity Perspective]{12 Years of Self-tracking for Promoting Physical Activity from a User Diversity Perspective: Taking Stock \& Thinking Ahead}

\author{Sofia Yfantidou}
\affiliation{%
  \institution{Aristotle University of Thessaloniki}
  \city{Thessaloniki}
  \country{Greece}
}
\email{syfantid@csd.auth.gr}
\orcid{0000-0002-5629-3493} 

\author{Pavlos Sermpezis}
\affiliation{%
  \institution{Aristotle University of Thessaloniki}
  \city{Thessaloniki}
  \country{Greece}
}
\email{sermpezis@csd.auth.gr}
\orcid{0000-0003-2129-977X}

\author{Athena Vakali}
\affiliation{%
  \institution{Aristotle University of Thessaloniki}
  \city{Thessaloniki}
  \country{Greece}
}
\email{avakali@csd.auth.gr}
\orcid{0000-0002-0666-6984} 

\renewcommand{\shortauthors}{Yfantidou, et al.}

\begin{abstract}
Despite the indisputable personal and societal benefits of regular physical activity, a large portion of the population does not follow the recommended guidelines, harming their health and wellness. The World Health Organization has called upon governments, practitioners, and researchers to accelerate action to address the global prevalence of physical inactivity. To this end, an emerging wave of research in ubiquitous computing has been exploring the potential of interactive self-tracking technology in encouraging positive health behavior change. Numerous findings indicate the benefits of personalization and inclusive design regarding increasing the motivational appeal and overall effectiveness of behavior change systems, with the ultimate goal of empowering and facilitating people to achieve their goals. However, most interventions still adopt a ``one-size-fits-all'' approach to their design, assuming equal effectiveness for all system features in spite of individual and collective user differences. To this end, we analyze a corpus of 12 years of research in self-tracking technology for health behavior change, focusing on physical activity, to identify those design elements that have proven most effective in inciting desirable behavior across diverse population segments. We then provide actionable recommendations for designing and evaluating behavior change self-tracking technology based on age, gender, occupation, fitness, and health condition. Finally, we engage in a critical commentary on the diversity of the domain and discuss ethical concerns surrounding tailored interventions and directions for moving forward.
\end{abstract}




\maketitle

\section{Introduction}
According to the World Health Organization (WHO), regular physical activity is amongst the determinants of good population health, generating significant personal and societal benefits. However, worldwide, 1 in 4 adults and 3 in 4 adolescents do not engage in regular physical activity, contributing to what has been coined as the ``inactivity pandemic'' \cite{world2019global}. To join the fight against the global prevalence of physical inactivity, technological innovations need to motivate individuals who are not sufficiently active to incorporate regular exercise into their daily routine. Research has shown that interactive technology can incite desirable health behavior change (HBC) in terms of healthy nutrition, regular physical activity, and disease management, among others \cite{roberts2017digital,lustria2013meta}. Ultimately though, physical activity interventions are the most prevalent among all HBC technological interventions, accounting for more than 1 out of 3 publications in the related literature \cite{Epstein2020}. 

% what is ST and how it can help users get more active
On the same note, emerging, ubiquitous self-tracking technology (ST) has been a game-changer for health and wellness promotion, focusing on physical activity. ST refers to ``the practice of gathering data about oneself on a regular basis and then recording and analysing the data to produce statistics and other data (such as images) relating to regular habits, behaviours and feelings'' \cite[\S1]{lupton2014self}. The philosophy behind it is that continuous monitoring can raise awareness regarding the user's behavioral patterns, whether beneficial or detrimental for their health, and assist them with improving their overall health outcomes. In recent years, a substantial amount of HCI research has been directed towards the design of such technologies, demonstrating the benefits and the persuasive power of strategically-designed ST for HBC \cite{Epstein2020}. However, despite the abundance of research studies in the field, prior work still has limitations in terms of system inclusive design, scattered domain knowledge, and technology abandonment. 

Specifically: (i) The \textbf{``one-size-fits-all'' mentality} is still prevalent in the majority of ST interventions' design; however, it might not always be fair across user segments. According to \citet{monteiro2019personalization}, interactive technology has not yet reached its full potential in terms of personalization and inclusive design, lacking context awareness, despite the great potential of personalized approaches in inciting positive HBC. (ii) There is \textbf{scattered knowledge concerning ST interventions adopting a diversity perspective}. It is unclear which kinds of personalization strategies exist, how to implement them in practice, and which of those strategies hold the most persuasive power for diverse user segments. In other words, given the large number of available system features (e.g., goal-setting, reminders) and related literature, it is unclear whether different features have different motivational appeal and, consequently, effectiveness in promoting an increase in physical activity or not. (iii) The adoption of ST might be widespread, but \textbf{technology abandonment is far from a rare phenomenon}. White papers report varied attrition rates for ST, ranging from 30\% to 70\% \cite{attrition1,attrition2}. While these numbers are not conclusive, they highlight the need for continuous evaluation of the user experience. However, there exist limited guidelines on personalizing the different performance metrics for diverse user groups.

% describe our work: a metadata analysis of PSD and past self
The limitations above highlight the importance of viewing the HBC domain through the lens of diversity and inclusive design, a perspective rarely explored in related literature. This paper aims to address this gap by conducting an exhaustive analysis and synthesis of prior knowledge accumulated in and expressed through two primary datasets (metadata analysis). While these data capture a comprehensive picture of the knowledge in the domain, they have never been analyzed through a diversity perspective, leading to the identification of personalization guidelines for HBC interventions. Specifically, our contributions are as follows:
\begin{description}
\item[C1 - Diversity analysis of ST feature effectiveness:]\hfill\\{We identify the most effective persuasive elements of ST, as per the Persuasive Systems Design (PSD) framework \cite{oinas2009persuasive}, based on the reported results of 117 prior interventions, totaling 6 million participants, stretching over 12 years (2008-2020). To achieve this, we conduct a previously unexplored personalized metadata analysis of the ``PAST SELF'' open corpus \cite{sofia_yfantidou_2020_4063377,yfantidou2021self}, where we compare the effectiveness of certain persuasive strategies and feature groups between diverse user segments, as well as the general population. Among others, we unveil that rewards barely hold any persuasive power for the elderly, opposite to the general population, or that social support strategies have a much stronger positive effect on gender-targeted interventions than mixed-gender interventions. We also identify unexplored dimensions in existing studies, which can serve as guidelines for future research in inclusive HBC interventions, focusing on physical activity. Through this contribution, we surface biases in ST design and move beyond the ``one-size-fits-all'' approach, comparing the persuasive power of ST features for diverse user segments by synthesizing metadata extracted from prior research.}
\item[C2 - Inclusive ST design \& evaluation recommendations: ]{We compare and contrast feature effectiveness and persuasive power among user groups of varying demographics, in terms of age, gender, occupation, fitness level, and health condition. The findings of this analysis reveal potential design and evaluation recommendations for fairer, more inclusive ST. Among others, our findings unveil the benefits of integrating human conversational agents or avatars in elderly-targeted ST interventions, given the high effectiveness of the ``Similarity'' persuasive strategy for this user segment and the exploitation of discussion forums and chatting functionality for gender-targeted interventions due to the high effectiveness of the ``Social Learning'' persuasive strategy in prior work. We accompany the aforementioned design recommendations with different sets of evaluation metrics that are better suited for certain feature and user groups. Following prior literature \cite{lalmas2014measuring}, we encourage a multi-faceted evaluation beyond quantitative data analysis by incorporating self-reported and qualitative metrics, as well as contextual factors and metadata, capturing multiple aspects of user engagement. 
% For example, we suggest the usage of system-user interaction metrics (e.g., feature access, absence time, step count between sessions) for the elderly users, and the usage of evaluation metrics capturing social interactions (e.g., content shared, time between content view and activity) for gender-targeted interventions. 
Through this contribution, we facilitate researchers and practitioners in designing and evaluating the ST user experience with the ultimate goal of providing inclusive functionality to avoid increased attrition rates.}
\end{description}

% structure
We structure the remaining of this paper as follows: Section~\ref{related-work} discusses HBC theories and their application for ST design. Section \ref{methodology} introduces the methodology of our metadata analysis, while Section \ref{results} presents and discusses our findings regarding feature effectiveness and evaluation for diverse user groups. Finally, Section \ref{conclusions} concludes our work and touches upon ethical implications and privacy concerns of personalization in personal informatics.

\section{Related Work\label{related-work}}
This section presents related work in the field of ST design for HBC. Section~\ref{rel:hbc} discusses behavior change theories within the ubiquitous computing domain, focusing on the PSD framework, which is crucial for our work. Section~\ref{rel:design_space} presents literature reviews and meta-analyses, revealing current challenges and the novelty of the presented work.

\subsection{Behavior Change Theories\label{rel:hbc}}
Ubiquitous computing and HCI researchers and practitioners have long understood the importance of theoretically founded design, designing, and implementing technological interventions inspired by numerous psychology, behavioral science, sports science, and behavioral economics theories. Behavior change theories can positively affect ST by informing design, guiding evaluation, and inspiring alternative experimental designs \cite{Hekler2013}. Some of the most widely-used theories in the domain \cite{yfantidou2021self} include the Social Cognitive Theory \cite{ratten2007social}, the Behaviour Change Technique Taxonomy \cite{michie2013behavior}, the Transtheoretical Model \cite{prochaska1997transtheoretical} and the Self-Determination Theory \cite{deci2008self}.

While behavior change theories provide indispensable high-level guidance, they do not describe how their theoretical concepts could be translated into real-world ST, leaving the interpretation to the respective researcher or practitioner. To bridge this gap, \citet{oinas2009persuasive}, among others, introduced the widely adopted PSD framework. PSD describes the content and functionality required in a behavior change product or service to increase its persuasive power. In particular, the PSD framework defines four persuasive strategies categories, i.e., \textit{primary task support}, \textit{dialogue support}, \textit{credibility}, and \textit{social support}; each having seven sub-groups within. However, personalization is out of the scope of the PSD framework, which adopts an ``one-size-fits-all'' approach. To address this limitation, we build upon the PSD taxonomy to conduct our metadata analysis, as described in Section~\ref{methodology}, from a user diversity perspective.

\subsection{ST Reviews and Meta-analyses: Design Space, Limitations \& Open Questions\label{rel:design_space}}
The exploitation of ubiquitous ST for increasing physical activity or decreasing sedentary behavior is an emerging field of study gathering growing scientific interest \cite{Epstein2020}. Thus, it is no surprise that numerous mapping and literature reviews and meta-analyses have attempted the synthesis of the aforementioned primary study results. Nevertheless, the majority do not adopt a diversity perspective.

A wave of research occupies itself with interventions targeting a specific user segment or PSD strategy. For instance, there exist reviews targeting users with specific traits, such as age \cite{elavsky2019mobile,quelly2016impact,Gerling2020}, race \cite{muller2016effectiveness}, occupation \cite{Babar2018b}, or physical and mental health conditions \cite{Murnane2018,spiel2019agency,nunes2015self}. Others assess the effectiveness of specific behavior change techniques or ST features, such as gamification \cite{hamari2014does}, chatting functionality \cite{buchholz2013physical}, or virtual and physical rewards \cite{strohacker2014impact}, social sharing \cite{Epstein2015c}, or the application of machine learning \cite{thieme2020machine} on the activity levels of individuals. However, contrary to the comprehensive nature of our work, the exact targeting of such works does not allow for comparative analysis between different population segments of persuasive strategies. 

Adopting a comparative approach, a number of studies consider broader inclusion and exclusion criteria for their primary studies, incorporating multiple user segments and functionalities. For instance, \citet{Epstein2020} published an exhaustive mapping review of the personal informatics literature, seeking to answer questions related to identifying areas of interest within personal informatics, tracking motivations, challenges, ethical concerns, and scientific contributions. While these reviews significantly contribute to the domain, their scope is orthogonal to our analysis, as they rely on empirical data to provide high-level guidelines. In contrast, our work is entirely data-driven, analyzing the outcomes of 12 years of related literature to unveil inclusive design and evaluation recommendations. On a similar note, \citet{matthews2016persuasive} and \citet{10.3389/fcomp.2020.00019} published systematic reviews assessing the effectiveness of mobile-application-based interventions in motivating physical activity identifying novel research directions. However, they neither assess the effectiveness of individual persuasive strategies nor personalize their recommendations to specific user groups. To address the first limitation, \citet{aldenaini2020trends} published a follow-up systematic review, evaluating the effectiveness of individual persuasive strategies in promoting physical activity. Similar to our work, they categorize intervention components under the PSD framework and report success rates per technique. However, they do not adopt a diversity perspective assuming equal effectiveness across diverse users. Closer to our work, \citet{yfantidou2021self} conducted a literature review of HBC ST interventions for increasing physical activity. Among others, they presented an assessment of the persuasive power of the PSD strategies based on the reported results of prior literature through the so-called ``PAST Score'', an effectiveness indicator. By exploiting the open-access corpus of this work \cite{sofia_yfantidou_2020_4063377}, we attempt the diversification of the ``PAST Score'', through our research methodology presented in Section~\ref{methodology}, and the presentation of guidelines for the design and evaluation of inclusive ST systems that are tailored to the needs and the goals of their users.

\section{Methodology\label{methodology}}
% metadata analysis: the studies we combine and whatever steps we need to report for such analysis
This section presents the methodology of our metadata analysis of two primary research studies from a user diversity perspective. Specifically, we utilize the open access ``PAST SELF'' corpus \cite{sofia_yfantidou_2020_4063377} and the PSD framework \cite{oinas2009persuasive} to assess ST feature effectiveness, and provide guidelines for personalized system design and evaluation.

\noindent{\textbf{The PSD Framework.}} As mentioned in Section~\ref{rel:hbc}, the PSD framework introduces 28 persuasion strategies grouped into four categories, namely, \textit{primary task support}, \textit{dialogue support}, \textit{credibility}, and \textit{social support}. PSD is a widely-adopted framework making it easier to compare our results with related meta-analyses and reviews \cite{aldenaini2020trends}, while it also enables us to compare and contrast primary studies integrating different theoretical elements by utilizing common constructs, i.e., the PSD strategies. We utilize an extended version of PSD to capture all persuasive strategies encountered in the studied literature. Specifically, we consider strategies, such as goal-setting, general information provision, punishment, and variability, which are not present in the original taxonomy (category ``Other'' from here onward). Goal-setting refers to system- or user-defined physical activity goals for the users to achieve; general information provision refers to sharing information with the users regards physical activity, health and general well-being; punishment refers to negative treatment for under-performance or failing to achieve preset goals; and variability refers to the system's ability to provide a variable experience to the user through variable rewards, game elements, interfaces, and hidden tasks. For definitions for the PSD framework's original strategies, see \cite{oinas2009persuasive}.

\noindent{\textbf{The ``PAST SELF'' Corpus.}} On a similar note, the ``PAST SELF'' Corpus is a rich source of metadata for 117 primary studies, containing information about the PSD strategies encountered in the included literature, as well as intervention characteristics and sample demographics for all studies (35 encoded fields in total). Note here that the corpus provides a complete overview of the literature in the domain for more than a decade. The corpus definition follows a systematic methodology to ensure the quality of included studies, based on the guidelines introduced by Kitchenham's \cite{kitchenham2007guidelines} widely recognized protocol for conducting systematic reviews. To locate the primary studies, the authors perform a broad search in Google Scholar, Scopus, IEEE Xplore, and Web of Science digital libraries utilizing a well-defined boolean search query fine-tuned according to the guidelines of \citet{spanos2016impact}. Overall, the authors screened 16774 articles after duplicate elimination, of which they removed 374 based on date criteria, 15112 based on the title, and 802 based on the abstract. After a full-text read of the remaining 546 articles, they excluded 429 articles based on predefined inclusion/exclusion criteria. Hence, 117 articles synthesized the final corpus. Inclusion criteria required a peer-review process, English language, a clearly-defined user intervention utilizing a ubiquitous device, and a quantitative assessment of the intervention's results. For our metadata analysis, we utilize the following subset of fields: 
\begin{itemize}
    \item Intervention Duration (in weeks): The intervention duration in number of weeks;
    \item Sample Size: The number of participants in the study; 
    \item Male: The number of male participants in the study;
    \item Special Criteria: Any special criteria of the participants, e.g., health condition;
    \item PSD - Primary Task Support: The Primary Task Support elements present in the intervention;
    \item PSD - Dialogue: The Dialogue elements present in the intervention;
    \item PSD - System Credibility: The System Credibility elements present in the intervention;
    \item PSD - Social Support: The Social Support elements present in the intervention;
    \item Independent Variable: All PSD strategies utilized in the intervention;
    \item Positive Result: A number from the set [-1, -0.5, 0, 0.5, 1], depending on the success of the intervention (-1 for statistically significant negative result and 1 for statistically significant positive result).
\end{itemize}
\begin{table}[ht!]
\vspace{-2mm}
\caption{Notation for the score generation functions of the personalized ``PAST Score''.\label{tab:notation}}
\vspace{-1mm}
\resizebox{\linewidth}{!}{%
\begin{tabular}{ll}
\hline
\multicolumn{1}{c}{{\textbf{Notation}}} & \multicolumn{1}{c}{{\textbf{Explanation}}} \\ \hline
$t_{i}$                                                                                  & Technique $i$, $t_i\in\{1,2,\ldots,32\}$                                                                            \\ \hline
$n_{papers}$                                                                                  & Total number of papers                                                                                \\ \hline
$p_{i,j_u}$                                                                              & Paper $j$ of intervention with user sample $u$ with technique $i$, 1 if technique $i$ appears in paper $j$, 0 otherwise                                                               \\ \hline
$r_{i,j_u}$                                                                                & Result of paper $j$ of intervention with user sample $u$ with technique $i$, $r_{i,j}\in\{-1, -0.5, 0, 0.5, 1\}$                                                                      \\ \hline
$w$                                                                                
& The preferred weight for the PAST\_score, $w\in[0,1]$                                                                       \\ \hline
\end{tabular}%
}
\vspace{-2mm}
\end{table}
% the calculation of the PAST score for each segment and the comparison with the general population

\noindent{\textbf{The Personalized PAST Score.}} Based on the encoded metadata, \cite{yfantidou2021self} calculates the so-called ``PAST Score'', an indicator of the effectiveness of each PSD strategy for the general population. However, we tweak the original formulas by introducing a personalization criterion moving beyond the ``one-size-fits-all'' approach. PAST score defines the efficacy and frequency of a technique $t_{i}$ (based on literature results) for the general population. We extend this, and define the efficacy of the technique $t_i$ for the user segment $u$, as the sum of the coded results of the $u$-sampled interventions the technique appears in, divided by the number of these papers, and the frequency of the technique $t_i$, as the number of $u$-sampled interventions that the technique appears divided by the total number of papers for this segment (for notations see Table~\ref{tab:notation}):
\begin{equation}
\resizebox{.4\textwidth}{!}{$\textrm{efficacy}(t_i,u)=\frac{\sum_{j_u} r_{i,{j_u}}}{\sum_{j_u} p_{i,{j_u}}}
\quad\text{and}\quad 
\textrm{frequency}(t_i,u)=\frac{\sum_{j_u} p_{i,{j_u}}}{\sum_{j_u} p_{j_u}}$}
\end{equation}
The scores are normalized in the $[+1,+5]$ interval for comparison purposes. Note here that each metric has its shortcomings, hence we report the metrics combined (i.e., PAST Score) but also separately in Section~\ref{results}. Specifically, efficacy alone can be misleading, since a rare strategy that appears in a single study with positive results would get a maximum score, ignoring its generalizability capacity. To  overcome this shortcoming, we also utilize a strategy's frequency, as an indicator of confidence in the strategy's persuasiveness capacity. However, frequency alone would punish rare but potentially ground-breaking results, while over-rewarding commonly used strategies. The final score for a technique is a combination of its reported personalized efficacy and usage frequency in the investigated papers. For our analysis, in equation \ref{eq:pastscore}, we assume a weight of $w=0.5$, i.e., equal importance for efficacy and frequency. Specifically:
\begin{equation}
\resizebox{.4\textwidth}{!}{$\textrm{Personalized\_PAST\_Score}(t_i,u)=w*\textrm{efficacy}(t_i,u)+(1-w)*\textrm{frequency}(t_i,u)$}
\label{eq:pastscore}
\end{equation}
% what user segments we explore/filter
Based on the personalized version of the ``PAST Score'', we then assess the effectiveness of different PSD strategies for diverse segments of the population. For each segment, we also compare PSD effectiveness with the general population. Additionally, we provide a comparison between large-scale interventions (sample size > 50 or intervention duration > 10 weeks) versus all included interventions to explore potential similarities and differences between them. Specifically, we choose to analyze population segments defined by: age (children, adolescents, elderly, mixed), gender (male, female), occupation (office workers, university students), fitness and physique (inactive population, overweight and obese population), and health condition (patients with cancer, diabetes type 2 or heart disease).

% Finally, we select evaluation metrics
\noindent{\textbf{The Evaluation Dimensions.}} To provide comprehensive guidance for HCI and ubiquitous computing researchers and practitioners, we accompany the PSD effectiveness with indicative features and evaluation metrics. We consider a spectrum of evaluation dimensions, measuring different qualities of the user experience, inspired by the work of \citet{lalmas2014measuring}. Based on the findings of our analysis, we recommend a combination of the following evaluation dimensions to capture the various aspects of user experience with ST. Specifically, we consider four distinct aspects: 
\begin{description}
    \vspace{-1mm}
    \item[Perceived Self Aspect:] the user's self-reported image of their everyday experiences, as well as psychological, technological, social, and health factors, usually measured through qualitative evaluation methods;
    \item[Physical Self Aspect:] the user's physical reaction to the interaction with the system, which can be interpreted as the physical activity performed in response to the system's intervention
    \item[Behavioral Self Aspect:] the user's behavioral response to the system, calculated via user-system interaction metrics, such as wear-time and session duration
    \item[Environmental Aspect:] the external factors that affect the user's interaction with the system and the execution of the desired behavior, such as weather or location.
\end{description}

\section{Analysis Results \& Discussion: Inclusive Feature Effectiveness \& Evaluation\label{results}}
This section discusses the most effective PSD strategies per population segment based on the personalized ``PAST Score''. It also unveils the unexplored ``corners'' of the ST design space for diverse segments, identifying research gaps to motivate future work in the field (Section~\ref{psd-general}). In the meantime, it provides a set of recommendations of indicative system features and matching evaluation metrics that are best suited for each use case and user segment (Section~\ref{use-cases}).

\subsection{PSD Strategies Effectiveness from a Personalized Perspective: Across Population Segments\label{psd-general}}
\begin{figure}[htb!]
    \centering
    \includegraphics[width=.9\linewidth]{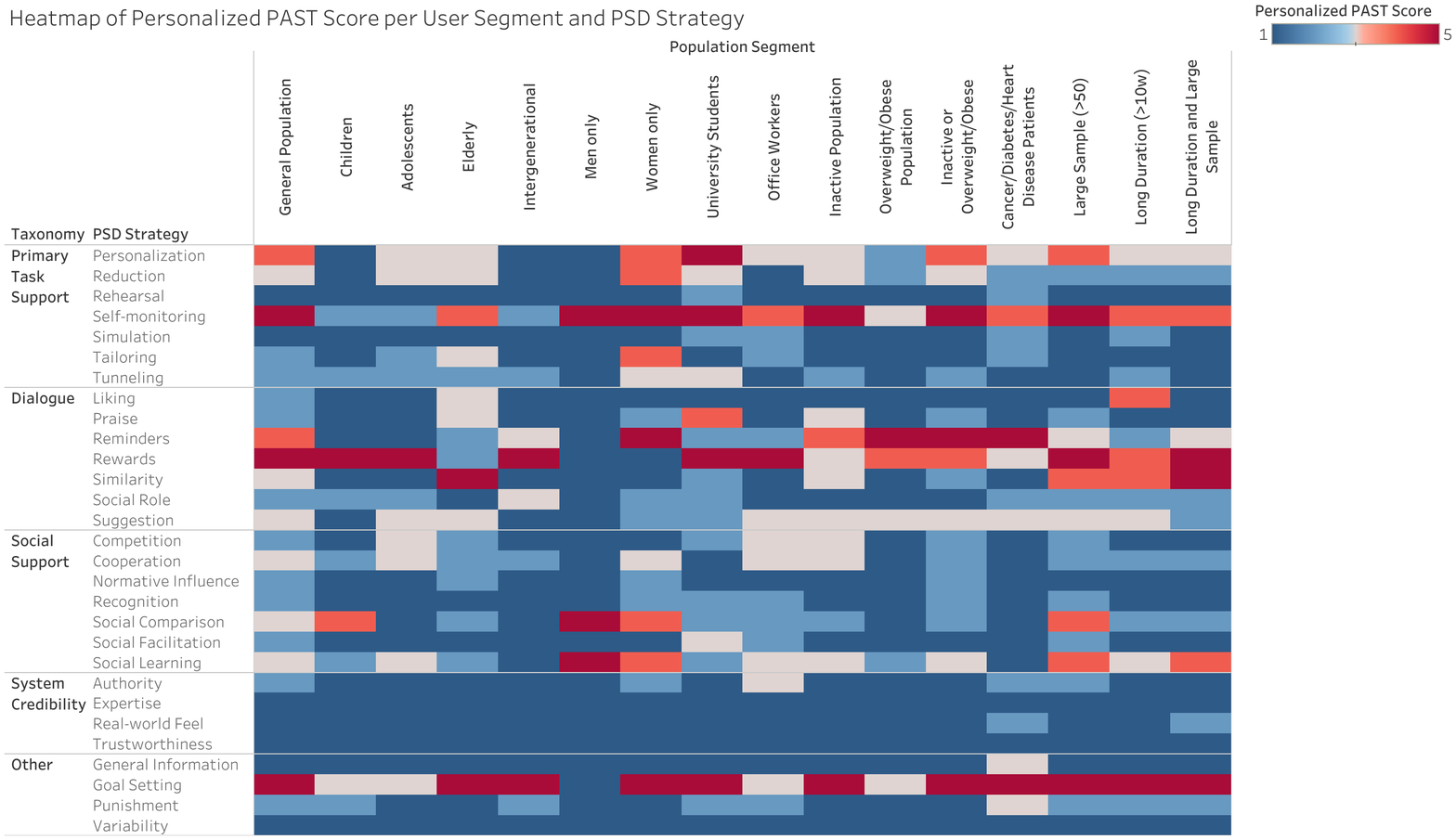}
    \vspace{-1cm}
    \caption{A heat map of persuasive strategies effectiveness (in terms of PAST Score) versus population segments. Blue color indicates lower effectiveness (1-2), while beige and red colors indicate medium (3) and high (4-5) effectiveness, respectively. The ``Other'' category indicates strategies of the extended PSD taxonomy.}
    \label{fig:heatmap}
    \vspace{-3mm}
\end{figure}

According to \cite{yfantidou2021self}, interventions that have utilized self-monitoring, goal-setting, and rewards tend to have higher success rates, indicating the effectiveness of such persuasive strategies for the general population. As a first extension, we explore the correlation between the extended PSD framework strategies' effectiveness (expressed through the personalized ``PAST Score'') and the different population segments (Figure~\ref{fig:heatmap}). Naturally, the aforementioned PSD strategies show higher effectiveness (red-like colors) across most population segments. Specifically, self-monitoring has a high or very high personalized ``PAST Score'' (+4 or +5) for 75\% of user segments and goal-setting and rewards for almost 70\%. In Section~\ref{use-cases} we will discuss possible exceptions to the rule. Overall, high scoring strategies (+4 and +5) are rare accounting for 14\% of the heat map, with medium (+3), and low or very low (+2 and +1) scoring strategies accounting for 13\% and 74\%, respectively. High-scoring strategies also tend to belong to certain categories. We notice that the ``Dialogue'' and ``Primary Task Support'' strategies show overall higher effectiveness ($\mu=2.33$, $\sigma=1.39$ and $\mu=2.13$, $\sigma=1.39$) compared to ``Other'' ($\mu=2.00$, $\sigma=1.51$), ``Social Support'' ($\mu=1.84$, $\sigma=0.98$) and ``System Credibility'' ($\mu=1.13$, $\sigma=0.38$) across multiple segments. 

It is important to note here that lower scores are not necessarily indicators of low effectiveness but also unexplored research domains. This is because the ``PAST Score'' incorporates a strategy's frequency into the scoring function; hence novel or unexplored strategies might appear with lower scores. While this might seem like a disadvantage at first, incorporating frequency is essential to avoid over-rewarding infrequent techniques for individual positive results. However, researchers and practitioners are encouraged to adapt the weight parameter, $w$, to their application's needs. 
% \begin{figure}[htb!]
%     \centering
%     \includegraphics[width=.8\linewidth]{img/Heatmap_frequency.pdf}
%     \vspace{-1cm}
%     \caption{A heat map of persuasive strategies frequency score versus population segments. Blue color indicates lower frequency (1-2), while beige and red colors indicate medium (3) and high (4-5) frequency, respectively.\label{fig:heatmap-freq}}
%     \vspace{-5mm}
% \end{figure}

\begin{figure}[htb!]
\centering
\begin{minipage}{.45\textwidth}
    \centering
    \includegraphics[width=\linewidth]{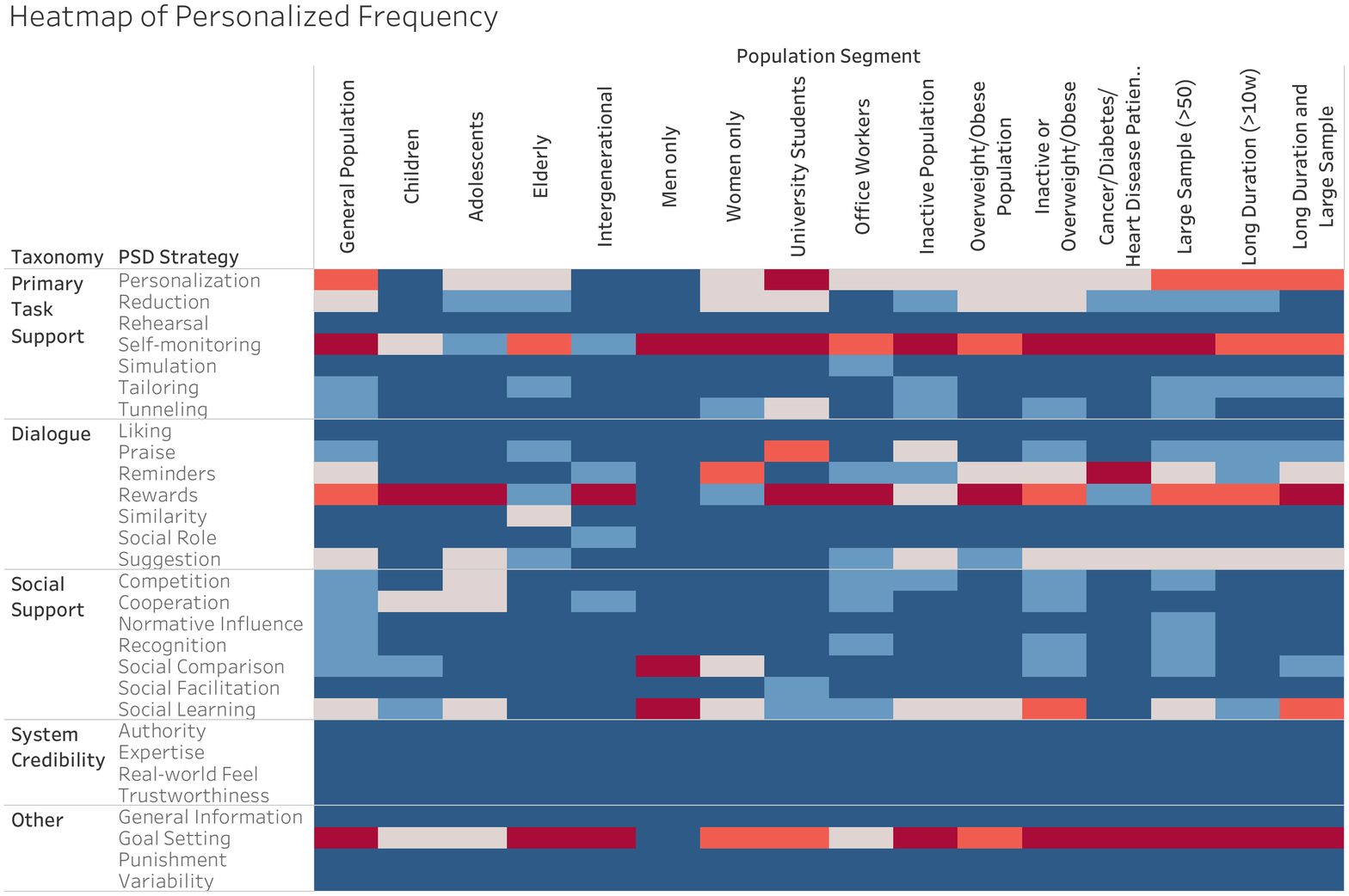}
    \caption{A heat map of persuasive strategies frequency score versus population segments.\label{fig:heatmap-freq}}
\end{minipage}%
\hspace{2mm}
\begin{minipage}{.45\textwidth}
    \centering
    \includegraphics[width=\linewidth]{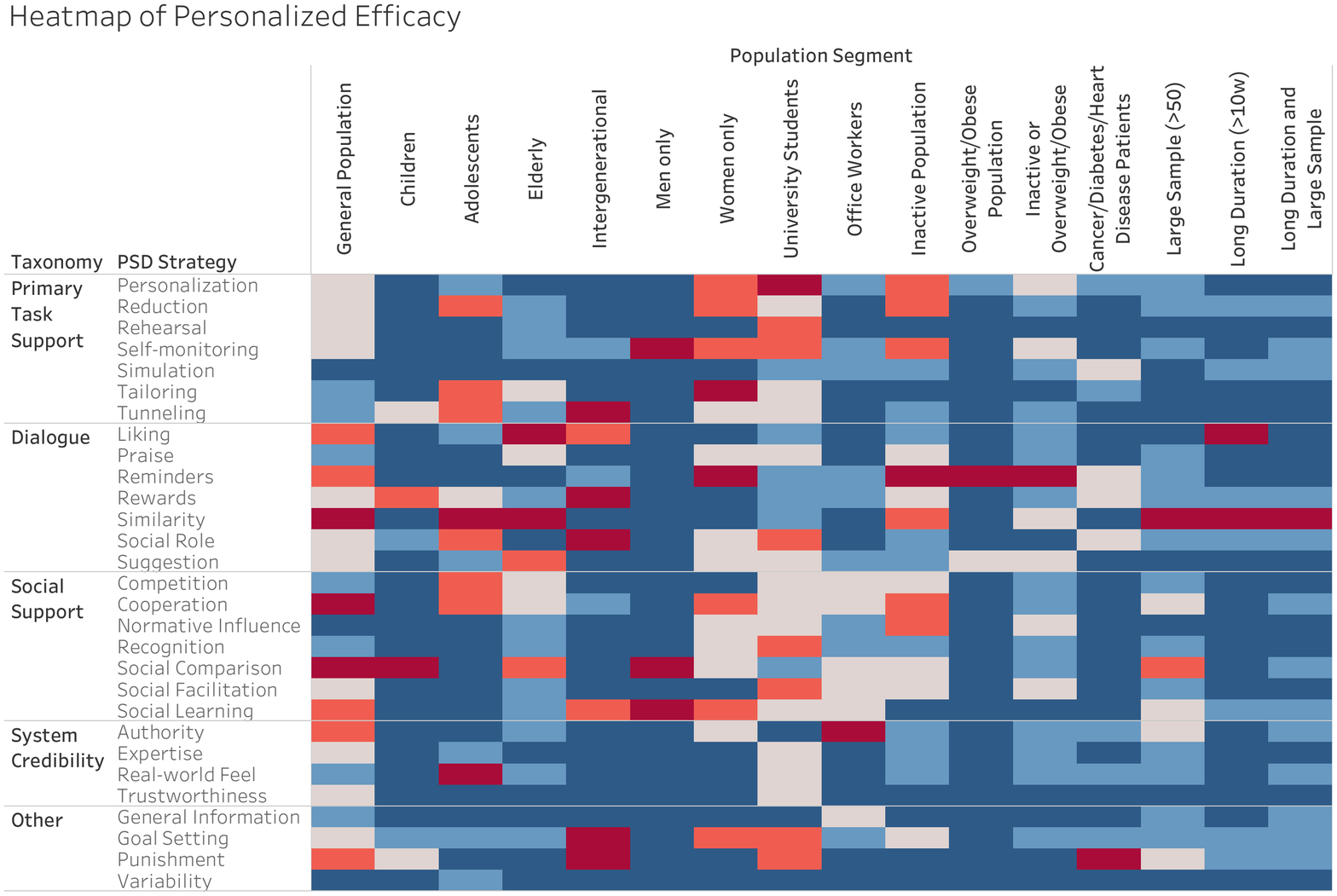}
    \caption{A heat map of persuasive strategies efficacy versus population segments..\label{fig:heatmap-eff}}
\end{minipage}
\vspace{-2mm}
\end{figure}
To this end, in Figure \ref{fig:heatmap-freq}, we present an alternative version of the heat map presenting the frequency scores of each PSD strategy across population segments ($w=0$). Immediately, we can see that all ``System Credibility'' strategies without exception have a meager frequency score (+1). Similarly, 3 out of 4 ``Other'' strategies (except for goal-setting) also have a very low frequency. Even in the popular categories, there are PSD strategies that have been barely experimented with. For instance, social facilitation (6\% of population segments above very low frequency), and normative influence (12\%) from the ``Social Support'' category, social role (6\%), similarity (6\%), and liking (0\%) from ``Dialogue'', and finally, simulation (6\%) and rehearsal (0\%) from ``Primary Task Support''. Note that a percentage of 0\% strategies above very low frequency does not necessarily indicate a full absence of related interventions but potentially minimal experimentation. Similarly, there exists an imbalance in the number of interventions exploiting various PSD strategies between different population segments. For example, women-targeted interventions are more common and diverse than men-targeted ones (31\% of PSD strategies above very low frequency vs. 10\%). Also, popular samples include inactive or overweight and obese populations (48\% of PSD strategies above very low frequency) and office workers (38\%). On the contrary, intergenerational and children-based interventions are rarer and more monotonous, as only 20\% of PSD strategies have above very low frequency score for these segments. Given this intuition, we can deduce that certain PSD strategies, such as those under ``System Credibility'' and ``Other'', or user segments, such as men and children (see note on ethics in Section~\ref{conclusions}) offer fertile ground for further exploration. On the contrary, other strategies, such as goal-setting and rewards, or user groups, such as women and office workers, have received ample scientific attention.

% \begin{figure}[htb!]
%     \centering
%     \includegraphics[width=.8\linewidth]{img/Heatmap_efficacy.pdf}
%     \vspace{-1cm}
%     \caption{A heat map of persuasive strategies efficacy versus population segments. Blue color indicates lower efficacy (1-2), while beige and red colors indicate medium (3) and high (4-5) efficacy, respectively.}
%     \label{fig:heatmap-eff}
%     \vspace{-3mm}
% \end{figure}
Figure \ref{fig:heatmap-eff} presents the opposite heat map, where we completely exclude the frequency factor from the personalized ``PAST Score'' ($w=1$), accounting only for reported efficacy. We notice that despite the significant lack of ``System Credibility'' strategies in the related literature, some show very high efficacy scores (+5) for certain population segments, e.g., real-world feel for adolescents or authority for office workers. This suggests that future self-tracking interventions incorporate these previously considered secondary features. Similarly, we notice that even techniques with low ``PAST Score'' for the general population show high or very high efficacy (+4 or +5) for certain segments, such as rehearsal for university students (+4 in efficacy compared to +1 generic ``PAST Score''), liking for the elderly, and intergenerational interventions (+5 and +4 from +2), or rehearsal for university students (+4 from +2). Such visualization provides us with incites about less tested but promising and potentially ground-breaking PSD strategies that pave the ground for future work in inclusive HBC interventions focusing on physical activity.

\begin{figure}[htb!]
    \centering
    \includegraphics[width=\linewidth]{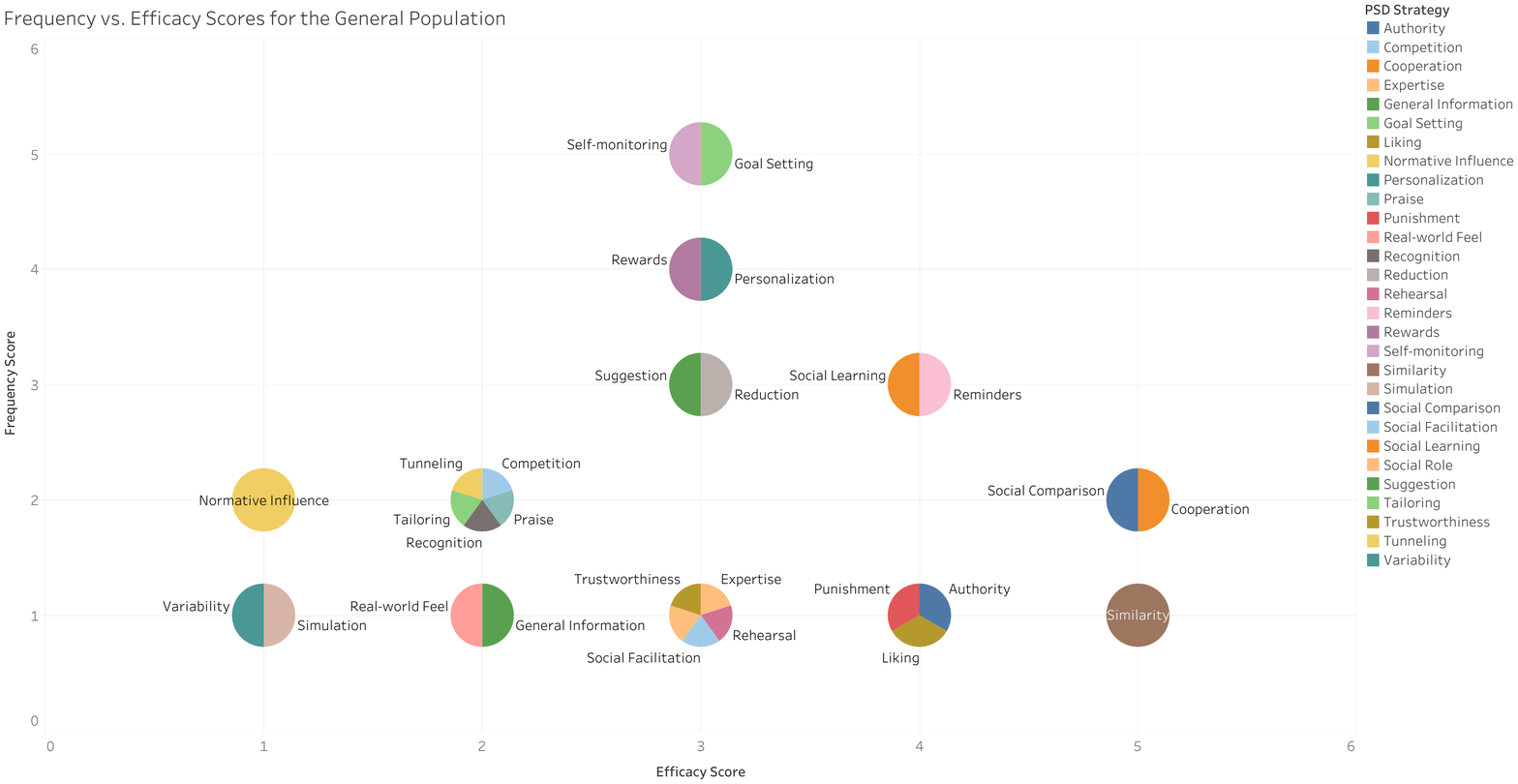}
    \vspace{-1.5cm}
    \caption{A scatter plot of efficacy scores (x-axis) versus frequency scores (y-axis) for each PSD strategy for the general population.}
    \label{fig:scatter}
    \vspace{-3mm}
\end{figure}
Summing up, Figure \ref{fig:scatter} gives us an overview of the comparison between PSD strategies' frequency (y-axis) versus efficacy (x-axis) in prior literature for the general population. Each circle in the figure represents one or more PSD strategies (overlapping strategies are shown as pie chart pieces). We notice that the most popular techniques, e.g., self-monitoring and goal-setting, ultimately show overall medium efficacy. At the same time, some popular techniques in commercial self-tracking devices, i.e., praise or competition, show overall low efficacy. Some others (see bottom left part of the plot) appear to be inefficient; however, this may be due to insufficient evidence, and we need more data to conclude their effectiveness. Nevertheless, the most interesting part of the plot -in terms of potential for future work- is the bottom-right corner, which presents the least tried techniques, i.e., similarity, social comparison, cooperation, liking, punishment, and authority, that have led to successful results in prior HBC interventions.

This section has given an overview of the persuasive power of the PSD strategies for distinct population segments while we further our discussion on a case-to-case basis below.

\vspace{-0.2cm}
\subsection{PSD Strategies Effectiveness from a Diversity Perspective: Use Cases\label{use-cases}}
This section presents a series of engaging, indicative use cases where we compare the effectiveness of each PSD strategy across diverse segments of the population through radar plots. PSD strategies are placed on the circle's circumference, whereas the inner polygons indicate the personalized ``PAST score'' for each strategy. The blue polygon refers to the general population, while the red (or green) to specific user segments.
\begin{figure}[htb!]
\vspace{-1mm}
\centering
\begin{minipage}{.45\textwidth}
    \centering
    \includegraphics[width=\linewidth]{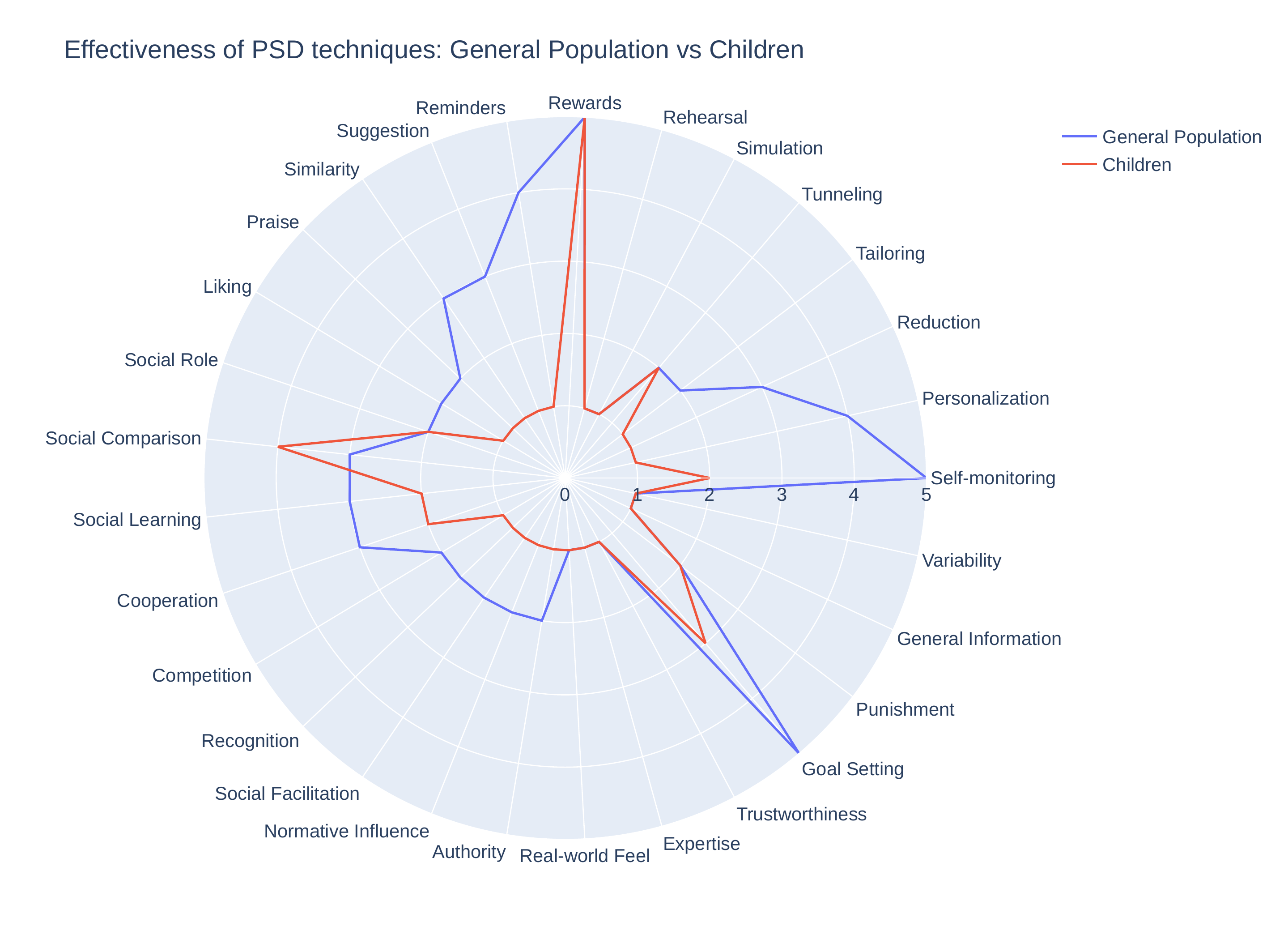}
    \caption{PSD Strategies effectiveness comparison between the general population and the children population.\label{fig:children}}
\end{minipage}%
\hspace{2mm}
\begin{minipage}{.45\textwidth}
    \centering
    \includegraphics[width=\linewidth]{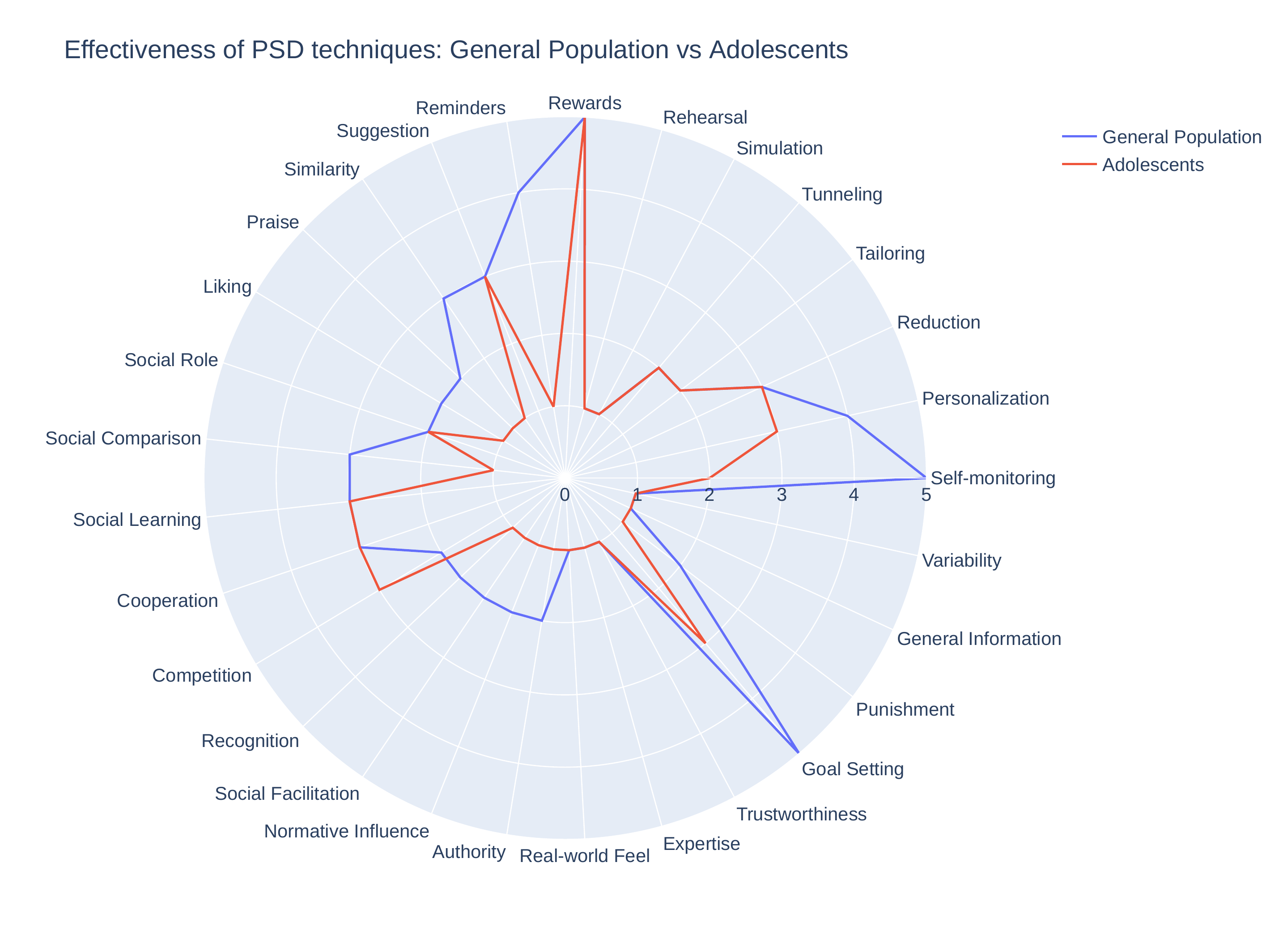}
    \caption{PSD Strategies effectiveness comparison between the general population and the adolescents population.\label{fig:adolescents}}
\end{minipage}
\vspace{-2mm}
\end{figure}

\subsubsection{Age-group Differences}
\begin{figure}[htb!]
\vspace{-1mm}
\centering
\begin{minipage}{.45\textwidth}
    \centering
    \includegraphics[width=.9\linewidth]{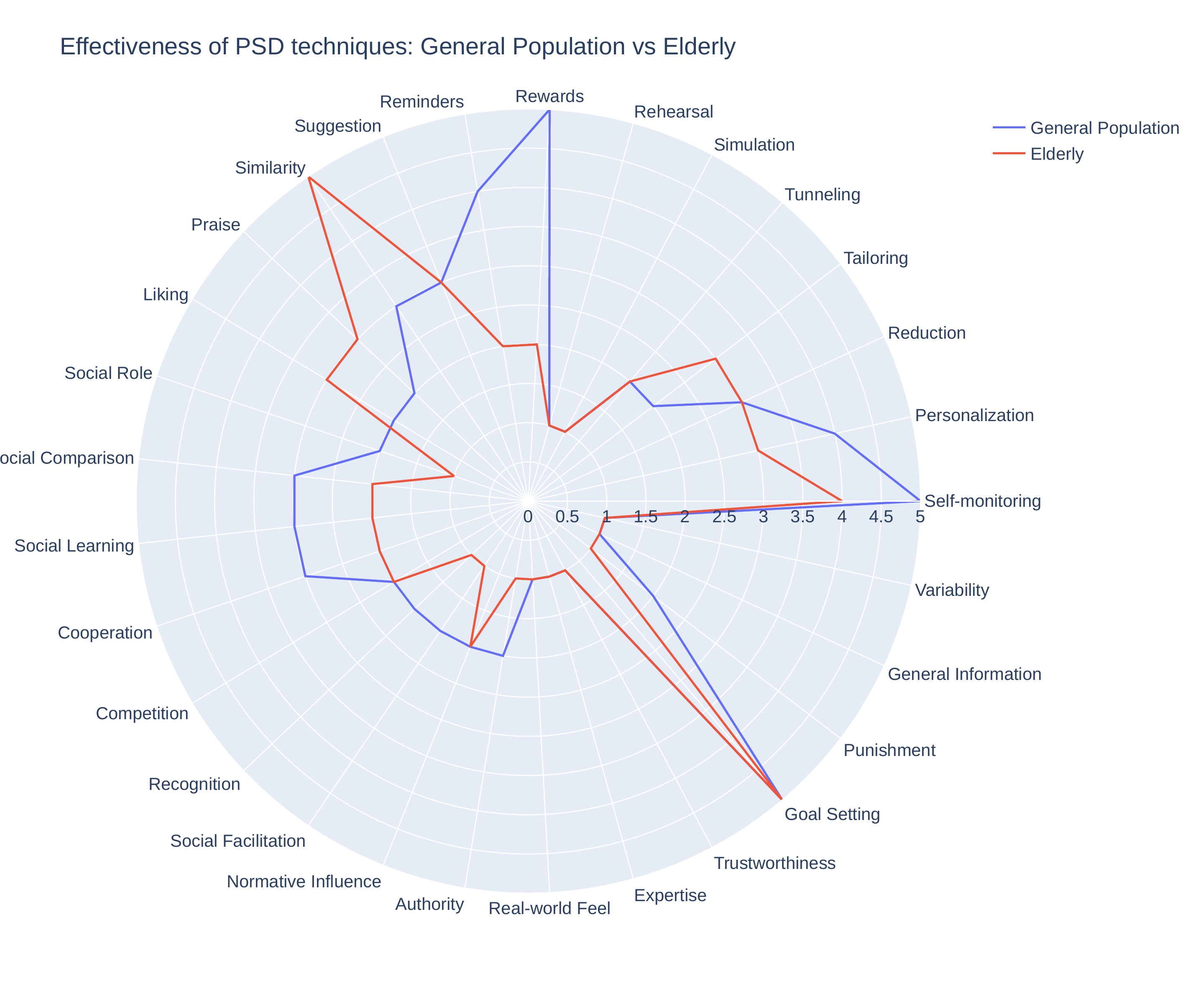}
    \caption{PSD Strategies effectiveness comparison between the general population and the elderly population.\label{fig:elderly}}
\end{minipage}%
\hspace{2mm}
\begin{minipage}{.45\textwidth}
    \centering
    \includegraphics[width=\linewidth]{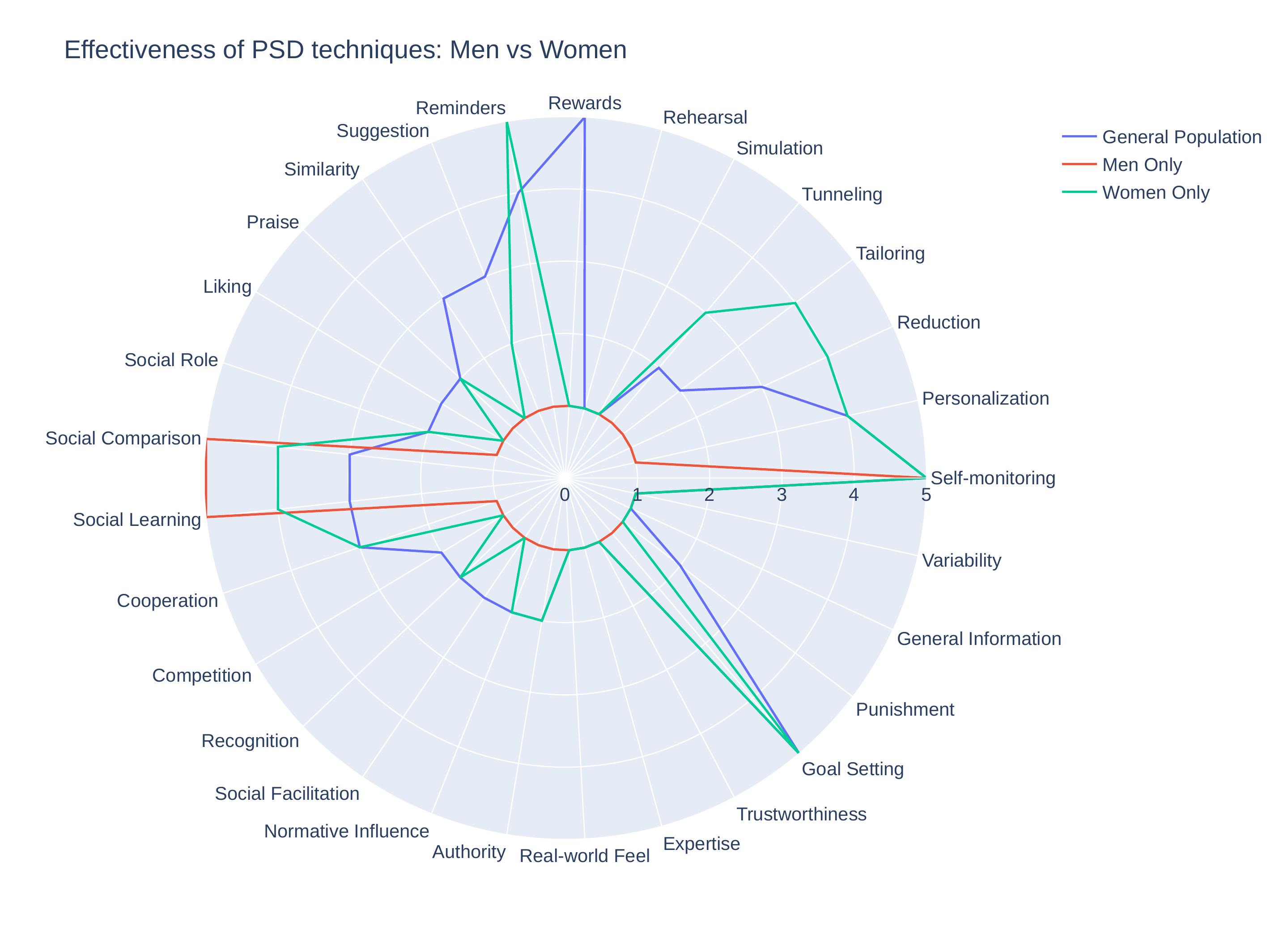}
    \caption{PSD Strategies effectiveness comparison between the general population, men and women.\label{fig:men_women}}
\end{minipage}
\vspace{-2mm}
\end{figure}

Initially, we explore differences in persuasive power across various age groups compared to the general population. The radar plots in Figures \ref{fig:children} and \ref{fig:adolescents} present a comparison of PSD strategies' effectiveness between young users (children and adolescents) and the general population, while the radar plot in Figure~\ref{fig:elderly} focuses on the elderly population. 

% \begin{figure}[htb!]
% \vspace{-2mm}
% \centering
% \begin{minipage}{.5\textwidth}
%   \centering
%   \includegraphics[width=.9\linewidth]{img/children.pdf}
% %   \captionof{figure}{Comparison between }
% %   \label{fig:test1}
% \end{minipage}%
% \begin{minipage}{.5\textwidth}
%   \centering
%   \includegraphics[width=.9\linewidth]{img/adolescents.pdf}
% %   \captionof{figure}{Another figure}
% %   \label{fig:test2}
% \end{minipage}
% \vspace{-1mm}
% \caption{PSD Strategies effectiveness comparison between the general population, children (left), and adolescents (right), respectively.\label{fig:children_adolescents}}
% \vspace{-3mm}
% \end{figure}

We notice that the most powerful strategy, based on reported results, for inciting positive HBC in children through ST is \textit{rewards} (+5), followed by \textit{social comparison} (+4), which shows more persuasive power for children compared to the general population. Rewards are usually tied to gamification in children-oriented interventions, where real-world physical activity is translated into virtual commodities. Examples of such rewards include game currency or extra playtime \cite{garde2015assessment,berkovsky2010physical}. With regards to social comparison, prior research has utilized decoy opponents used for real-life comparisons \cite{berkovsky2010physical}. On the contrary, there is a large drop in ``Primary Task Support'' effectiveness (e.g., self-monitoring +2 vs. +5) and goal-setting (+3 vs. +5). We conjecture that such features, designed and fine-tuned on adult populations, might not be directly applicable to children. Hence, further research is required towards designing kid-friendly primary task support for positive HBC. Given the complexity of self-reporting, we could recommend a combination of indicative physical and interaction factors (Physical Self Aspect, Behavioral Self Aspect; see Section \ref{methodology}) for evaluating the effectiveness of features in interventions with children populations. Specifically, the adoption of desired behavior can be measured through tracked data, such as activity duration, or type of activities performed. At the same time, the interaction success can be evaluated via feature access, number of sessions, activity performed between sessions, or number of acquired commodities and rewards.

% \begin{tcolorbox}
% \textbf{CHILDREN.}\\
% \textbf{Recommended strategies:} Virtual Rewards, Social Comparison\\
% \textbf{Evaluation metrics:} Activity between sessions, Acquired rewards, Age-based activity duration comparison\\
% \textbf{Strategies needing further exploration:} Self-monitoring, Goal-setting, Personalization
% \end{tcolorbox}
\vspace{0.3cm}
\noindent\fbox{%
    \parbox{\columnwidth}{%
        \textbf{CHILDREN.}\\
        \textbf{Recommended strategies:} Virtual Rewards, Social Comparison\\
        \textbf{Evaluation metrics:} Activity between sessions, Acquired rewards, Age-based activity duration comparison\\
        \textbf{Strategies needing further exploration:} Self-monitoring, Goal-setting, Personalization
    }%
}
\vspace{0.1cm}

Similarly, interventions on adolescents (Figure~\ref{fig:adolescents}) utilizing \textit{rewards} also show high effectiveness (+5), but ``Social Support'' features, such as \textit{social learning}, \textit{cooperation} and \textit{competition} offer more promising results compared to Social Comparison (+3 vs. +1). Rewards (material or virtual) in adolescent interventions slightly diverge from gamification practices. Material rewards take the form of monetary prizes and gifts, while virtual rewards move away from in-game commodities to badges and point systems \cite{corepal2019feasibility,blackman2015examining,klausen2016effects,mendoza2017fitbit}. Additionally, discussion groups in social media platforms have been integrated in HBC interventions as a means of social facilitation \cite{mendoza2017fitbit,corepal2019feasibility}, similarly to individual and group-based competitions \cite{corepal2019feasibility,blackman2015examining}. Note that the effectiveness of ``Primary Task Support'' strategies, such as personalization and reduction, seems to gradually increase with age. Evaluation metrics here should capture the social nature of adolescent-based interventions. Apart from tracked data, e.g., activity duration, time between sessions, energy expenditure (Physical Self Aspect), it is important to collect interaction metrics (Behavioral Self Aspect), such as content views (e.g., messages, posts, e-mails), notifications, and content response time, content shares, comments, and likes, as well as user content generation. Similarly, environmental factors (Environmental Aspect) such as social network and interactions can offer a more comprehensive view of the intervention's success.

\vspace{0.3cm}
% \begin{tcolorbox}
% \textbf{ADOLESCENTS.}\\
% \textbf{Recommended strategies:} Virtual and Physical Rewards, Social Facilitation, Cooperation, Competition\\
% \textbf{Evaluation metrics:} User content generation, Comments, Shares, Interactions, User Network\\
% \textbf{Strategies needing further exploration:} Self-monitoring, Goal-setting, Personalization
% \end{tcolorbox}
\noindent\fbox{%
    \parbox{\columnwidth}{%
        \textbf{ADOLESCENTS.}\\
        \textbf{Recommended strategies:} Virtual and Physical Rewards, Social Facilitation, Cooperation, Competition\\
        \textbf{Evaluation metrics:} User content generation, Comments, Shares, Interactions, User Network\\
        \textbf{Strategies needing further exploration:} Self-monitoring, Goal-setting, Personalization
    }%
}
\vspace{0.1cm}
\begin{figure}[htb!]
\centering
\begin{minipage}{.45\textwidth}
  \centering
  \includegraphics[width=\linewidth]{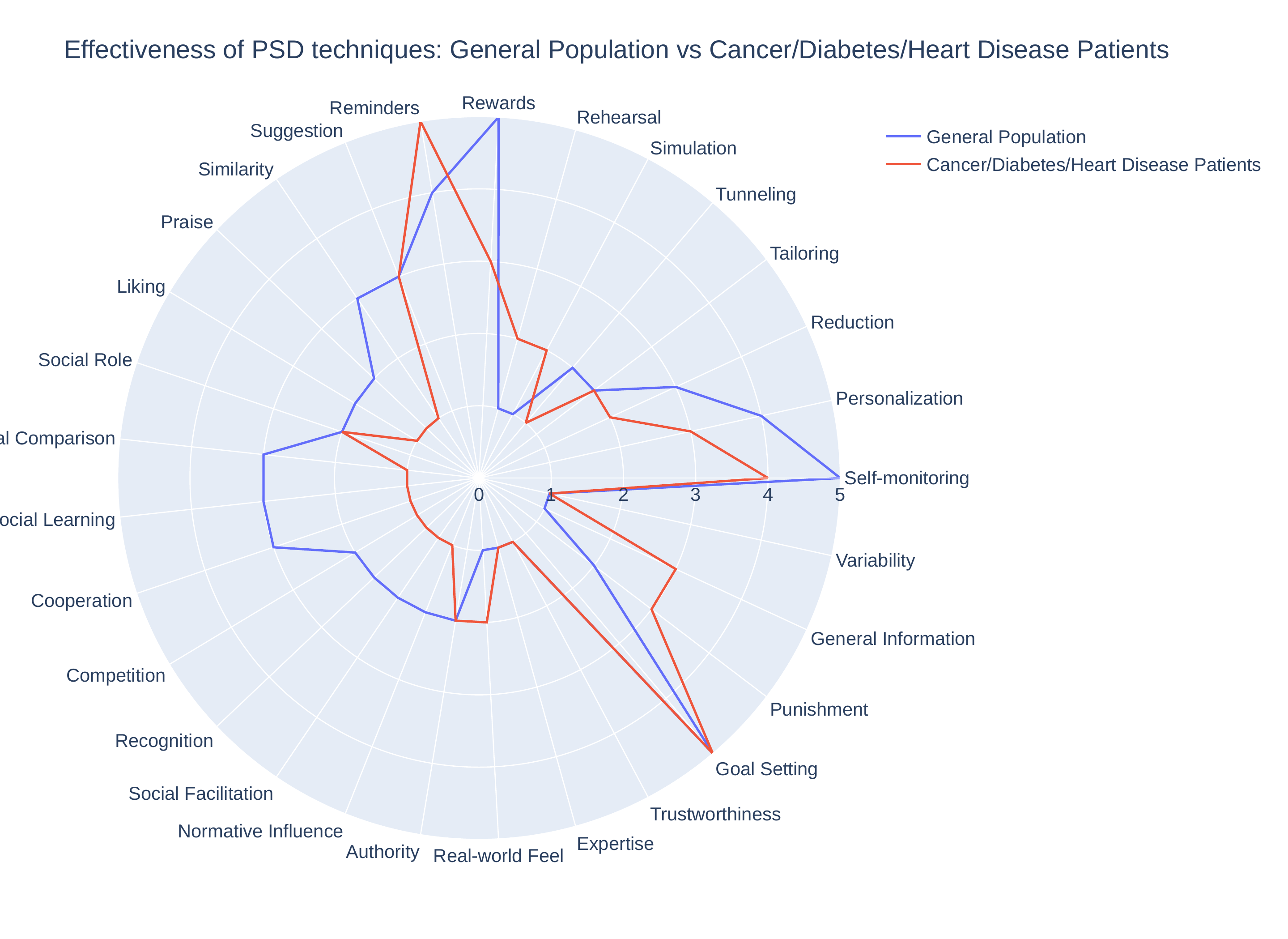}
  \captionof{figure}{PSD Strategies effectiveness comparison between the general population, and cancer, diabetes, and heart disease patients.}
  \label{fig:patients}
\end{minipage}%
\hspace{2mm}
\begin{minipage}{.45\textwidth}
  \centering
  \includegraphics[width=\linewidth]{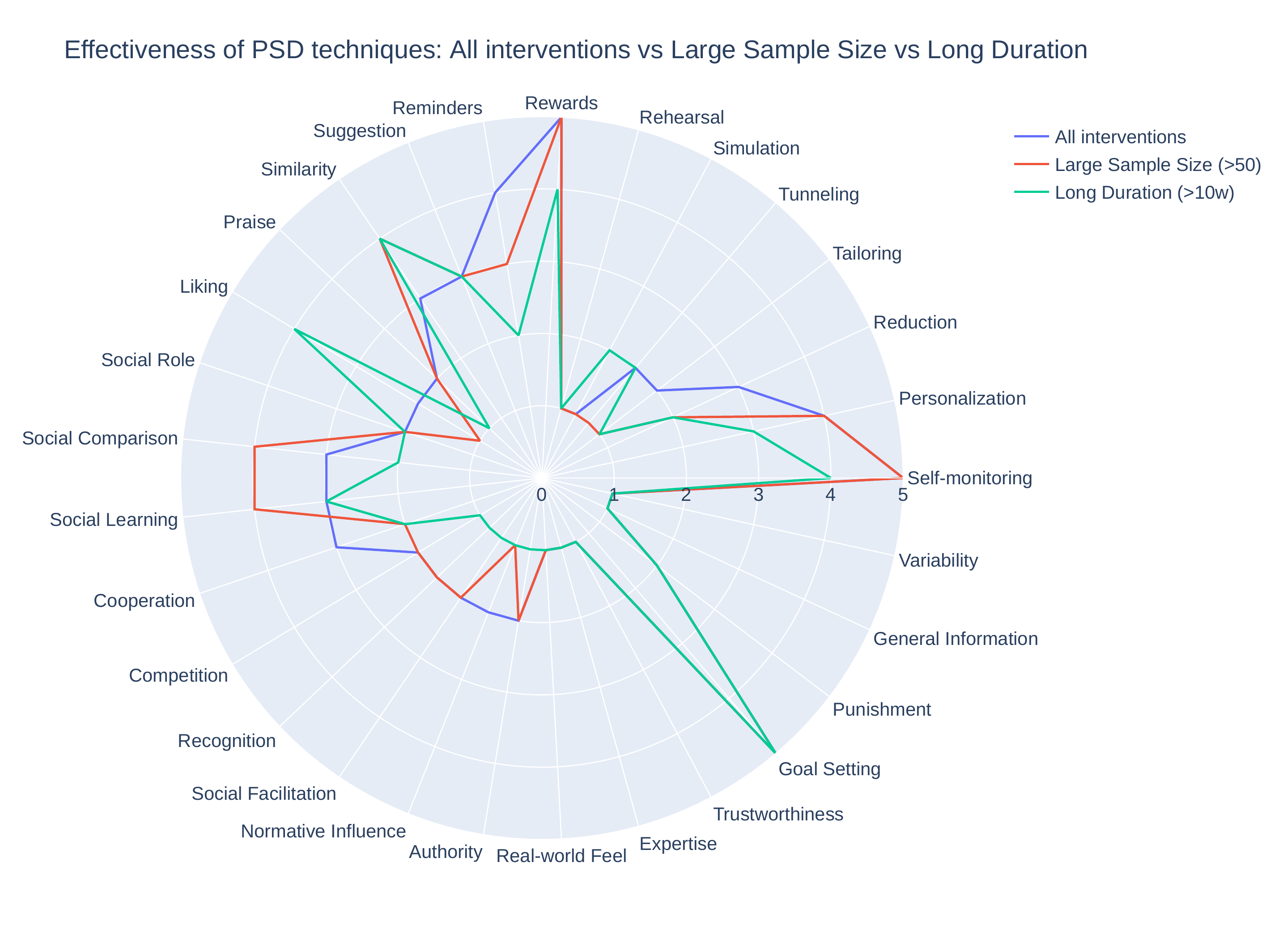}
  \captionof{figure}{PSD Strategies effectiveness comparison between all interventions, and interventions with a larger sample (>50 users) or a longer duration (>10 weeks).}
  \label{fig:large_long}
\end{minipage}
\vspace{-2mm}
\end{figure}

When it comes to the elderly population (Figure~\ref{fig:elderly}), the radar plot shows a different picture. ``Social Support'' strategies, i.e., social comparison, social learning, and cooperation, show decreased effectiveness (+1 or +2), giving ground to ``Dialogue'' strategies, such as \textit{liking}, \textit{praise} (+3) and especially \textit{similarity} (+5). Similarity usually takes the form of embodied conversational agents and human avatars \cite{bickmore2013randomized,alsaqer2017helping}. Liking is implemented via voice-controlled personal assistants \cite{kim2013text}, while praise can range from simple positive feedback for goal attainment to virtual pets exhibiting positive or negative emotions depending on the user behavior \cite{bickmore2013randomized}. The common denominator amongst these features is that they provide a sense of human connection to elderly users who might experience solitude. On the contrary, rewards show a significant decrease in effectiveness for this population segment, contrary to younger samples. For this segment of the population, relevant evaluation metrics include sedentariness duration, or type of activities performed, measuring the direct effect of the intervention (Physical Self Aspect), as well as feature access, content views, interactions with conversational agents, and wear time (Behavioral Self Aspect). Also, self-reported physical and mental health or physical activity competency are relevant factors when tailoring interventions for older users (Perceived Self Aspect).

\vspace{0.3cm}
% \begin{tcolorbox}
% \textbf{ELDERLY.}\\
% \textbf{Recommended strategies:} Liking, Praise, Similarity\\
% \textbf{Evaluation metrics:} Sedentariness duration, feature access, conversational agent interactions, health\\
% \textbf{Strategies needing further exploration:} Rewards, Tailoring
% \end{tcolorbox}
\noindent\fbox{%
    \parbox{\columnwidth}{%
        \textbf{ELDERLY.}\\
        \textbf{Recommended strategies:} Liking, Praise, Similarity\\
        \textbf{Evaluation metrics:} Sedentariness duration, feature access, conversational agent interactions, health\\
        \textbf{Strategies needing further exploration:} Rewards, Tailoring 
    }%
}

\subsubsection{Gender Differences}
With regards to gender differences, Figure~\ref{fig:men_women} presents a comparison of the effectiveness of PSD strategies for men and women versus the general population. We notice that ``Social Support'' strategies, i.e., \textit{social comparison} and \textit{social learning}, show higher effectiveness for gender-specific interventions compared to generic interventions, especially for male samples (+5 vs. +4 for females vs. +3 for general population). Also, rewards have surprisingly low scores (+1) compared to the general population. However, as seen from the limited red spikes in the radar plot, interventions targeted to males are rare in the related literature, offering grounds for future work. 

With regards to female-based interventions, mainly \textit{tailoring} (+2 increment), and \textit{tunneling}, \textit{reduction}, and \textit{reminders} to a smaller extend (+1 increment), show promising results with regards to effectiveness, exceeding that of the general population. Reduction can be translated into system features through goal-setting guidance and progress bars for supporting partial goal accomplishment \cite{arigo2015addressing,zhang2019mobile}. Tailoring can take the form of contextualized suggestions for short active breaks \cite{finkelstein2015mobile}, and tunneling has been implemented as information provision accompanied by data-driven suggestions \cite{david2012walking,arigo2015addressing}. With regards to social features, gender-based interventions utilize more closed social platforms than adolescent-based interventions, such as private communities or team chats to create a collective sense of trust for progress sharing \cite{arigo2015addressing,zhang2019mobile}. Given the high importance of social features across genders, evaluation metrics should incorporate social factors, such as self-reported social support, group cohesion, social comparison tendencies (Perceived Self Aspect), as well as interaction analytics, such as content shares and views, and time between content view and activity (Behavioral Self Aspect). At the same time, evaluation metrics targeted to ``Primary Task Support'' strategies are also suitable, especially for female-based interventions, given their high effectiveness for this segment. Specifically, reduction, tailoring, and tunneling feature access should be evaluated by the activity performed as a result (Physical Self Aspect).

% \vspace{0.3cm}
% \begin{tcolorbox}
% \textbf{FEMALES.}\\
% \textbf{Recommended strategies:} Goal-setting, Reduction, Tailoring, Social Learning, Social Comparison, Reminders \\
% \textbf{Evaluation metrics:} Self-reported social support, Group cohesion, Social comparison tendencies, Interaction Analytics, Feature Access, Time between access and physical activity\\
% \textbf{Strategies needing further exploration:} Rewards
% \end{tcolorbox}
\vspace{0.3cm}
\noindent\fbox{%
    \parbox{\columnwidth}{%
        \textbf{FEMALES.}\\
        \textbf{Recommended strategies:} Goal-setting, Reduction, Tailoring, Social Learning, Social Comparison, Reminders \\
        \textbf{Evaluation metrics:} Self-reported social support, Group cohesion, Social comparison tendencies, Interaction Analytics, Feature Access, Time between access and physical activity\\
        \textbf{Strategies needing further exploration:} Rewards
    }%
}

% \begin{tcolorbox}
% \textbf{MALES.}\\
% \textbf{Recommended strategies:} Self-monitoring, Social Learning, Social Comparison \\
% \textbf{Evaluation metrics:} Social support, Group cohesion, Social comparison tendencies, Interaction Analytics\\
% \textbf{Strategies needing further exploration:} Rewards, Goal-setting, Reminders, Personalization
% \end{tcolorbox}
\noindent\fbox{%
    \parbox{\columnwidth}{%
        \textbf{MALES.}\\
        \textbf{Recommended strategies:} Self-monitoring, Social Learning, Social Comparison \\
        \textbf{Evaluation metrics:} Social support, Group cohesion, Social comparison tendencies, Interaction Analytics\\
        \textbf{Strategies needing further exploration:} Rewards, Goal-setting, Reminders, Personalization
    }%
}

\subsubsection{Health-related Differences}
Concerning patients suffering from a physical health condition, such as diabetes type 2, heart disease, or cancer (Figure~\ref{fig:patients}), we directly notice that ``Social Support'' strategies have a lower ``PAST Score'' compared to the general population, while \textit{reminders} and \textit{goal-setting} share the highest ``PAST Score'' (+5). Low scores, though, can be partly attributed to the low frequency of these PSD strategies for this segment (See Figure~\ref{fig:heatmap-freq}), indicating a direction of future work. On the other hand, \textit{reminders}, \textit{punishment}, \textit{real-world feel}, \textit{rehearsal}, and \textit{simulation} show a slight increase in effectiveness compared to the general population (+1), while provision of \textit{general information} shows the largest increase (+2). We assume that this is because this segment has a higher need for information about which forms of physical activity are suitable for their condition. Punishment can take the form of monetary punishment for failure to reach goals or negative feedback \cite{kim2018stretcharms,chokshi2018loss}, while real-world feel can refer to app store responsiveness \cite{ciravegna2019active} or contact form availability for support \cite{kooiman2018self}. With regards to primary-task support, rehearsal has been implemented as short exercise demonstration videos \cite{klausen2016effects}, and simulation as a linkage between real-world physical activity and virtual world experience \cite{hochsmann2019effectiveness}. To evaluate the user experience for this segment, one needs to incorporate self-reported metrics of health condition, and goals and expectations (Perceived Self Aspect), tracked metrics (Physical Self Aspect), as well as interaction metrics, such as content views (Behavioral Self Aspect) to evaluate for example information utility.
% \begin{tcolorbox}
% \textbf{CANCER/DIABETES/HEART DISEASE PATIENTS.}\\
% \textbf{Recommended strategies:} Goal-setting, Reminders, Information provision, Punishment \\
% \textbf{Evaluation metrics:} Health condition, Goals \& Expectations, Goal accomplishment rate, Content views\\
% \textbf{Strategies needing further exploration:} Social comparison, Social learning, Cooperation
% \end{tcolorbox}

\vspace{0.3cm}
\noindent\fbox{%
    \parbox{\columnwidth}{%
        \textbf{CANCER/DIABETES/HEART DISEASE PATIENTS.}\\
        \textbf{Recommended strategies:} Goal-setting, Reminders, Information provision, Punishment \\
        \textbf{Evaluation metrics:} Health condition, Goals \& Expectations, Goal accomplishment rate, Content views\\
        \textbf{Strategies needing further exploration:} Social comparison, Social learning, Cooperation
    }%
}
\subsubsection{Observed Differences based on Sample Size and Intervention Duration}
This section presents the reported results of large-scale and long-term interventions to indicate the longevity of the PSD strategies' effectiveness. We define large-scale interventions as those with a sample size larger than 50 participants and long-term interventions as those with a duration longer than 10 weeks. Amongst the first things we can notice is that \textit{goal-setting} sustains its maximum ``PAST Score'' both for large-scale and long-term interventions, whereas \textit{self-monitoring}, and \textit{rewards}' effectiveness whines with time (+4 from +5). This is in accordance with prior literature \cite{kramer2019cluster}, which reports diminished long-term effectiveness of incentives for inciting and maintaining HBC. On the contrary, ``Social Support'' features, such as \textit{social comparison} and \textit{social learning}, show improved effectiveness for large-scale interventions (+4 from +3). Another interesting point is that despite similarity and liking being quite rare in the included literature, they are correlated with better effectiveness (+1 and +2 increment, respectively in ``PAST Score'') for long-term interventions.

\section{Conclusion\label{conclusions}}
This paper presents a metadata analysis and related discussion performed on top of the PSD framework and an open corpus of 117 HBC and ST interventions. Through our exploration, we debunk the ``one-size-fits-all'' mentality in the design space of personal informatics, highlighting the significance of inclusive design, diversity and personalization. Specifically, we analyze the effectiveness of various persuasive strategies across diverse population segments and showcase the differences and similarities between them and the general population (C1). Additionally, we accompany the analysis with data-driven recommendations on system features and evaluation metrics suitable for each segment (C2) and provide directions for future work. We also provide an online interface for a more interactive experience\footnote{\url{https://syfantid.github.io/past-framework-visualization}}.

While we focus on physical activity interventions, future work may validate our results in different HBC domains, such as smoking cessation, or diet monitoring, to investigate possible changes in the perceived effectiveness of PSD strategies. It is also important to note that while design guidelines can facilitate the development of ST, researchers and practitioners need to be mindful of their epistemic status. While we strongly believe in the value of empirical data for generating design guidelines, we suggest that our recommendations be treated similarly to ``design hypotheses'', which require additional testing. As a final note, our results illustrate a lack of research on how different population segments (e.g., based on age, gender, and health status) interact with ST, which presents an opportunity for future work. For instance, the limited number of interventions targeting racial minorities did not allow us to perform an analysis based on race. Future research should focus on equitable access to ST for HBC, taking into account the socioeconomic status, the variable health, and the technological literacy of diverse population segments while being sensitive to their cultural and linguistic needs. On a similar note, privacy was rarely touched upon in our corpus. However, data privacy (e.g., data sharing, retention policy, third party involvement) is fundamental for trusted and secure use of ST, especially for sensitive groups such as minors or minorities, whose data sharing might raise ethical concerns. 

\begin{acks}
This project has received funding from the European Union’s Horizon 2020 research and innovation programme under the Marie Skłodowska-Curie grant agreement No 813162. The content of this paper reflects only the authors' view and the Agency and the Commission are not responsible for any use that may be made of the information it contains. The authors would also like to thank the web developers, T. Valk and S. Karamanidis, for their contribution to the development of the online exploration tool.
\end{acks}

\bibliographystyle{ACM-Reference-Format}
\bibliography{sample-base}


\begin{thebibliography}{54}

%%% the Web address.

\ifx \showCODEN    \undefined \def \showCODEN     #1{\unskip}     \fi
\ifx \showDOI      \undefined \def \showDOI       #1{#1}\fi
\ifx \showISBNx    \undefined \def \showISBNx     #1{\unskip}     \fi
\ifx \showISBNxiii \undefined \def \showISBNxiii  #1{\unskip}     \fi
\ifx \showISSN     \undefined \def \showISSN      #1{\unskip}     \fi
\ifx \showLCCN     \undefined \def \showLCCN      #1{\unskip}     \fi
\ifx \shownote     \undefined \def \shownote      #1{#1}          \fi
\ifx \showarticletitle \undefined \def \showarticletitle #1{#1}   \fi
\ifx \showURL      \undefined \def \showURL       {\relax}        \fi
% invisible to TeX
\providecommand\bibfield[2]{#2}
\providecommand\bibinfo[2]{#2}
\providecommand\natexlab[1]{#1}
\providecommand\showeprint[2][]{arXiv:#2}

\bibitem[Ger(2020)]%
        {Gerling2020}
 \bibinfo{year}{2020}\natexlab{}.
\newblock \showarticletitle{{Critical Reflections on Technology to Support
  Physical Activity among Older Adults}}.
\newblock \bibinfo{journal}{\emph{ACM Transactions on Accessible Computing}}
  \bibinfo{volume}{13}, \bibinfo{number}{1} (\bibinfo{year}{2020}).
\newblock
\showISSN{19367228}
\urldef\tempurl%
\url{https://doi.org/10.1145/3374660}
\showDOI{\tempurl}


\bibitem[Aldenaini et~al\mbox{.}(2020b)]%
        {10.3389/fcomp.2020.00019}
\bibfield{author}{\bibinfo{person}{Noora Aldenaini}, \bibinfo{person}{Oladapo
  Oyebode}, \bibinfo{person}{Rita Orji}, {and} \bibinfo{person}{Srinivas
  Sampalli}.} \bibinfo{year}{2020}\natexlab{b}.
\newblock \showarticletitle{Mobile Phone-Based Persuasive Technology for
  Physical Activity and Sedentary Behavior: A Systematic Review}.
\newblock \bibinfo{journal}{\emph{Frontiers in Computer Science}}
  \bibinfo{volume}{2} (\bibinfo{year}{2020}), \bibinfo{pages}{19}.
\newblock
\showISSN{2624-9898}
\urldef\tempurl%
\url{https://doi.org/10.3389/fcomp.2020.00019}
\showDOI{\tempurl}


\bibitem[Aldenaini et~al\mbox{.}(2020a)]%
        {aldenaini2020trends}
\bibfield{author}{\bibinfo{person}{Noora~Fahad Aldenaini},
  \bibinfo{person}{Felwah Alqahtani}, \bibinfo{person}{Rita Orji}, {and}
  \bibinfo{person}{Srinivas Sampalli}.} \bibinfo{year}{2020}\natexlab{a}.
\newblock \showarticletitle{Trends in Persuasive Technologies for Physical
  Activity and Sedentary Behavior: A Systematic Review}.
\newblock \bibinfo{journal}{\emph{Frontiers in Artificial Intelligence}}
  \bibinfo{volume}{3} (\bibinfo{year}{2020}), \bibinfo{pages}{7}.
\newblock


\bibitem[Alsaqer and Chatterjee(2017)]%
        {alsaqer2017helping}
\bibfield{author}{\bibinfo{person}{Mohammed Alsaqer} {and}
  \bibinfo{person}{Samir Chatterjee}.} \bibinfo{year}{2017}\natexlab{}.
\newblock \showarticletitle{Helping the elderly with physical exercise:
  Development of persuasive mobile intervention sensitive to elderly cognitive
  decline}. In \bibinfo{booktitle}{\emph{2017 IEEE 19th International
  Conference on e-Health Networking, Applications and Services (Healthcom)}}.
  IEEE, \bibinfo{pages}{1--6}.
\newblock


\bibitem[Arigo et~al\mbox{.}(2015)]%
        {arigo2015addressing}
\bibfield{author}{\bibinfo{person}{Danielle Arigo}, \bibinfo{person}{Leah~M
  Schumacher}, \bibinfo{person}{Emilie Pinkasavage}, {and}
  \bibinfo{person}{Meghan~L Butryn}.} \bibinfo{year}{2015}\natexlab{}.
\newblock \showarticletitle{Addressing barriers to physical activity among
  women: A feasibility study using social networking-enabled technology}.
\newblock \bibinfo{journal}{\emph{Digital health}}  \bibinfo{volume}{1}
  (\bibinfo{year}{2015}), \bibinfo{pages}{2055207615583564}.
\newblock


\bibitem[Babar et~al\mbox{.}(2018)]%
        {Babar2018b}
\bibfield{author}{\bibinfo{person}{Yash Babar}, \bibinfo{person}{Jason Chan},
  {and} \bibinfo{person}{Ben Choi}.} \bibinfo{year}{2018}\natexlab{}.
\newblock \showarticletitle{{“Run forrest run!”: Measuring the impact of
  App-enabled performance and social feedback on running performance}}. In
  \bibinfo{booktitle}{\emph{International Conference on Information Systems
  2018, ICIS 2018}}, Vol.~\bibinfo{volume}{2023}.
  \bibinfo{publisher}{Association for Information Systems (AIS)},
  \bibinfo{address}{San Francisco, California, United States},
  \bibinfo{pages}{1--17}.
\newblock
\showISBNx{9780996683173}


\bibitem[Berkovsky et~al\mbox{.}(2010)]%
        {berkovsky2010physical}
\bibfield{author}{\bibinfo{person}{Shlomo Berkovsky}, \bibinfo{person}{Mac
  Coombe}, \bibinfo{person}{Jill Freyne}, \bibinfo{person}{Dipak Bhandari},
  {and} \bibinfo{person}{Nilufar Baghaei}.} \bibinfo{year}{2010}\natexlab{}.
\newblock \showarticletitle{Physical activity motivating games: virtual rewards
  for real activity}. In \bibinfo{booktitle}{\emph{Proceedings of the SIGCHI
  Conference on Human Factors in Computing Systems}}.
  \bibinfo{pages}{243--252}.
\newblock


\bibitem[Bickmore et~al\mbox{.}(2013)]%
        {bickmore2013randomized}
\bibfield{author}{\bibinfo{person}{Timothy~W Bickmore},
  \bibinfo{person}{Rebecca~A Silliman}, \bibinfo{person}{Kerrie Nelson},
  \bibinfo{person}{Debbie~M Cheng}, \bibinfo{person}{Michael Winter},
  \bibinfo{person}{Lori Henault}, {and} \bibinfo{person}{Michael~K
  Paasche-Orlow}.} \bibinfo{year}{2013}\natexlab{}.
\newblock \showarticletitle{A randomized controlled trial of an automated
  exercise coach for older adults}.
\newblock \bibinfo{journal}{\emph{Journal of the American Geriatrics Society}}
  \bibinfo{volume}{61}, \bibinfo{number}{10} (\bibinfo{year}{2013}),
  \bibinfo{pages}{1676--1683}.
\newblock


\bibitem[Blackman et~al\mbox{.}(2015)]%
        {blackman2015examining}
\bibfield{author}{\bibinfo{person}{Kacie~CA Blackman}, \bibinfo{person}{Jamie
  Zoellner}, \bibinfo{person}{Adil Kadir}, \bibinfo{person}{Brandon Dockery},
  \bibinfo{person}{Sallie~Beth Johnson}, \bibinfo{person}{Fabio~A Almeida},
  \bibinfo{person}{D~Scott McCrickard}, \bibinfo{person}{Jennie~L Hill},
  \bibinfo{person}{Wen You}, {and} \bibinfo{person}{Paul~A Estabrooks}.}
  \bibinfo{year}{2015}\natexlab{}.
\newblock \showarticletitle{Examining the feasibility of smartphone game
  applications for physical activity promotion in middle school students}.
\newblock \bibinfo{journal}{\emph{Games for health journal}}
  \bibinfo{volume}{4}, \bibinfo{number}{5} (\bibinfo{year}{2015}),
  \bibinfo{pages}{409--419}.
\newblock


\bibitem[Buchholz et~al\mbox{.}(2013)]%
        {buchholz2013physical}
\bibfield{author}{\bibinfo{person}{Susan~Weber Buchholz},
  \bibinfo{person}{Joellen Wilbur}, \bibinfo{person}{Diana Ingram}, {and}
  \bibinfo{person}{Louis Fogg}.} \bibinfo{year}{2013}\natexlab{}.
\newblock \showarticletitle{Physical activity text messaging interventions in
  adults: a systematic review}.
\newblock \bibinfo{journal}{\emph{Worldviews on Evidence-Based Nursing}}
  \bibinfo{volume}{10}, \bibinfo{number}{3} (\bibinfo{year}{2013}),
  \bibinfo{pages}{163--173}.
\newblock


\bibitem[Chokshi et~al\mbox{.}(2018)]%
        {chokshi2018loss}
\bibfield{author}{\bibinfo{person}{Neel~P Chokshi}, \bibinfo{person}{Srinath
  Adusumalli}, \bibinfo{person}{Dylan~S Small}, \bibinfo{person}{Alexander
  Morris}, \bibinfo{person}{Jordyn Feingold}, \bibinfo{person}{Yoonhee~P Ha},
  \bibinfo{person}{Marta~D Lynch}, \bibinfo{person}{Charles~AL Rareshide},
  \bibinfo{person}{Victoria Hilbert}, {and} \bibinfo{person}{Mitesh~S Patel}.}
  \bibinfo{year}{2018}\natexlab{}.
\newblock \showarticletitle{Loss-framed financial incentives and personalized
  goal-setting to increase physical activity among ischemic heart disease
  patients using wearable devices: the ACTIVE REWARD randomized trial}.
\newblock \bibinfo{journal}{\emph{Journal of the American Heart Association}}
  \bibinfo{volume}{7}, \bibinfo{number}{12} (\bibinfo{year}{2018}),
  \bibinfo{pages}{e009173}.
\newblock


\bibitem[Ciravegna et~al\mbox{.}(2019)]%
        {ciravegna2019active}
\bibfield{author}{\bibinfo{person}{Fabio Ciravegna}, \bibinfo{person}{Jie Gao},
  \bibinfo{person}{Neil Ireson}, \bibinfo{person}{Robert Copeland},
  \bibinfo{person}{Joe Walsh}, {and} \bibinfo{person}{Vitaveska Lanfranchi}.}
  \bibinfo{year}{2019}\natexlab{}.
\newblock \showarticletitle{Active 10: Brisk walking to support regular
  physical activity}. In \bibinfo{booktitle}{\emph{Proceedings of the 13th EAI
  International Conference on pervasive computing technologies for
  healthcare}}. \bibinfo{pages}{11--20}.
\newblock


\bibitem[Corepal et~al\mbox{.}(2019)]%
        {corepal2019feasibility}
\bibfield{author}{\bibinfo{person}{Rekesh Corepal}, \bibinfo{person}{Paul
  Best}, \bibinfo{person}{Roisin O’Neill}, \bibinfo{person}{Frank Kee},
  \bibinfo{person}{Jennifer Badham}, \bibinfo{person}{Laura Dunne},
  \bibinfo{person}{Sarah Miller}, \bibinfo{person}{Paul Connolly},
  \bibinfo{person}{Margaret~E Cupples}, \bibinfo{person}{Esther~MF Van~Sluijs},
  {et~al\mbox{.}}} \bibinfo{year}{2019}\natexlab{}.
\newblock \showarticletitle{A feasibility study of ‘The StepSmart
  Challenge’to promote physical activity in adolescents}.
\newblock \bibinfo{journal}{\emph{Pilot and feasibility studies}}
  \bibinfo{volume}{5}, \bibinfo{number}{1} (\bibinfo{year}{2019}),
  \bibinfo{pages}{1--15}.
\newblock


\bibitem[David et~al\mbox{.}(2012)]%
        {david2012walking}
\bibfield{author}{\bibinfo{person}{Prabu David}, \bibinfo{person}{Janet
  Buckworth}, \bibinfo{person}{Michael~L Pennell}, \bibinfo{person}{Mira~L
  Katz}, \bibinfo{person}{Cecilia~R DeGraffinreid}, {and}
  \bibinfo{person}{Electra~D Paskett}.} \bibinfo{year}{2012}\natexlab{}.
\newblock \showarticletitle{A walking intervention for postmenopausal women
  using mobile phones and interactive voice response}.
\newblock \bibinfo{journal}{\emph{Journal of telemedicine and telecare}}
  \bibinfo{volume}{18}, \bibinfo{number}{1} (\bibinfo{year}{2012}),
  \bibinfo{pages}{20--25}.
\newblock


\bibitem[Deci and Ryan(2008)]%
        {deci2008self}
\bibfield{author}{\bibinfo{person}{Edward~L Deci} {and}
  \bibinfo{person}{Richard~M Ryan}.} \bibinfo{year}{2008}\natexlab{}.
\newblock \showarticletitle{Self-determination theory: A macrotheory of human
  motivation, development, and health.}
\newblock \bibinfo{journal}{\emph{Canadian psychology/Psychologie canadienne}}
  \bibinfo{volume}{49}, \bibinfo{number}{3} (\bibinfo{year}{2008}),
  \bibinfo{pages}{182}.
\newblock


\bibitem[Elavsky et~al\mbox{.}(2019)]%
        {elavsky2019mobile}
\bibfield{author}{\bibinfo{person}{Steriani Elavsky}, \bibinfo{person}{Lenka
  Knapova}, \bibinfo{person}{Adam Klocek}, {and} \bibinfo{person}{David
  Smahel}.} \bibinfo{year}{2019}\natexlab{}.
\newblock \showarticletitle{Mobile health interventions for physical activity,
  sedentary behavior, and sleep in adults aged 50 years and older: a systematic
  literature review}.
\newblock \bibinfo{journal}{\emph{Journal of aging and physical activity}}
  \bibinfo{volume}{27}, \bibinfo{number}{4} (\bibinfo{year}{2019}),
  \bibinfo{pages}{565--593}.
\newblock


\bibitem[Epstein et~al\mbox{.}(2020)]%
        {Epstein2020}
\bibfield{author}{\bibinfo{person}{Daniel~A. Epstein}, \bibinfo{person}{Clara
  Caldeira}, \bibinfo{person}{Mayara~Costa Figueiredo}, \bibinfo{person}{Xi
  Lu}, \bibinfo{person}{Lucas~M. Silva}, \bibinfo{person}{Lucretia Williams},
  \bibinfo{person}{Jong~Ho Lee}, \bibinfo{person}{Qingyang Li},
  \bibinfo{person}{Simran Ahuja}, \bibinfo{person}{Qiuer Chen},
  \bibinfo{person}{Payam Dowlatyari}, \bibinfo{person}{Craig Hilby},
  \bibinfo{person}{Sazeda Sultana}, \bibinfo{person}{Elizabeth~V. Eikey}, {and}
  \bibinfo{person}{Yunan Chen}.} \bibinfo{year}{2020}\natexlab{}.
\newblock \showarticletitle{{Mapping and Taking Stock of the Personal
  Informatics Literature}}.
\newblock \bibinfo{journal}{\emph{Proceedings of the ACM on Interactive,
  Mobile, Wearable and Ubiquitous Technologies}} \bibinfo{volume}{4},
  \bibinfo{number}{4} (\bibinfo{year}{2020}).
\newblock
\showISSN{24749567}
\urldef\tempurl%
\url{https://doi.org/10.1145/3432231}
\showDOI{\tempurl}


\bibitem[Epstein et~al\mbox{.}(2015)]%
        {Epstein2015c}
\bibfield{author}{\bibinfo{person}{Daniel~A. Epstein},
  \bibinfo{person}{Bradley~H. Jacobson}, \bibinfo{person}{Elizabeth Bales},
  \bibinfo{person}{David~W. McDonald}, {and} \bibinfo{person}{Sean~A. Munson}.}
  \bibinfo{year}{2015}\natexlab{}.
\newblock \showarticletitle{{From "nobody cares" to "way to go!": A design
  framework for social sharing in personal informatics}}.
\newblock \bibinfo{journal}{\emph{CSCW 2015 - Proceedings of the 2015 ACM
  International Conference on Computer-Supported Cooperative Work and Social
  Computing}} (\bibinfo{year}{2015}), \bibinfo{pages}{1622--1636}.
\newblock
\showISBNx{9781450329224}
\urldef\tempurl%
\url{https://doi.org/10.1145/2675133.2675135}
\showDOI{\tempurl}


\bibitem[Finkelstein et~al\mbox{.}(2015)]%
        {finkelstein2015mobile}
\bibfield{author}{\bibinfo{person}{Joseph Finkelstein},
  \bibinfo{person}{McKenzie Bedra}, \bibinfo{person}{Xuan Li},
  \bibinfo{person}{Jeffrey Wood}, {and} \bibinfo{person}{Pamela Ouyang}.}
  \bibinfo{year}{2015}\natexlab{}.
\newblock \showarticletitle{Mobile App to Reduce Inactivity in Sedentary
  Overweight Women.}. In \bibinfo{booktitle}{\emph{MedInfo}}.
  \bibinfo{pages}{89--92}.
\newblock


\bibitem[Gandhi(2015)]%
        {attrition2}
\bibfield{author}{\bibinfo{person}{Malay Gandhi}.}
  \bibinfo{year}{2015}\natexlab{}.
\newblock \bibinfo{booktitle}{\emph{Deconstructing the Fitbit IPO and S-1}}.
\newblock Rock Health.
\newblock
\urldef\tempurl%
\url{https://rockhealth.com/deconstructing-fitbit-s-1/}
\showURL{%
Retrieved July 28, 2020 from \tempurl}


\bibitem[Garde et~al\mbox{.}(2015)]%
        {garde2015assessment}
\bibfield{author}{\bibinfo{person}{Ainara Garde}, \bibinfo{person}{Aryannah
  Umedaly}, \bibinfo{person}{S~Mazdak Abulnaga}, \bibinfo{person}{Leah
  Robertson}, \bibinfo{person}{Anne Junker}, \bibinfo{person}{Jean~Pierre
  Chanoine}, \bibinfo{person}{J~Mark Ansermino}, {and} \bibinfo{person}{Guy~A
  Dumont}.} \bibinfo{year}{2015}\natexlab{}.
\newblock \showarticletitle{Assessment of a mobile game (“MobileKids Monster
  Manor”) to promote physical activity among children}.
\newblock \bibinfo{journal}{\emph{Games for health journal}}
  \bibinfo{volume}{4}, \bibinfo{number}{2} (\bibinfo{year}{2015}),
  \bibinfo{pages}{149--158}.
\newblock


\bibitem[Hamari et~al\mbox{.}(2014)]%
        {hamari2014does}
\bibfield{author}{\bibinfo{person}{Juho Hamari}, \bibinfo{person}{Jonna
  Koivisto}, {and} \bibinfo{person}{Harri Sarsa}.}
  \bibinfo{year}{2014}\natexlab{}.
\newblock \showarticletitle{Does gamification work?--a literature review of
  empirical studies on gamification}. In \bibinfo{booktitle}{\emph{2014 47th
  Hawaii international conference on system sciences}}.
  \bibinfo{publisher}{IEEE}, \bibinfo{address}{Washington, DC, United States},
  \bibinfo{pages}{3025--3034}.
\newblock


\bibitem[Hekler et~al\mbox{.}(2013)]%
        {Hekler2013}
\bibfield{author}{\bibinfo{person}{Eric~B. Hekler}, \bibinfo{person}{Predrag
  Klasnja}, \bibinfo{person}{Jon~E. Froehlich}, {and}
  \bibinfo{person}{Matthew~P. Buman}.} \bibinfo{year}{2013}\natexlab{}.
\newblock \showarticletitle{{Mind the theoretical gap: Interpreting, using, and
  developing behavioral theory in HCI research}}.
\newblock \bibinfo{journal}{\emph{Conference on Human Factors in Computing
  Systems - Proceedings}} (\bibinfo{year}{2013}), \bibinfo{pages}{3307--3316}.
\newblock
\showISBNx{9781450318990}
\urldef\tempurl%
\url{https://doi.org/10.1145/2470654.2466452}
\showDOI{\tempurl}


\bibitem[H{\"o}chsmann et~al\mbox{.}(2019)]%
        {hochsmann2019effectiveness}
\bibfield{author}{\bibinfo{person}{Christoph H{\"o}chsmann},
  \bibinfo{person}{Denis Infanger}, \bibinfo{person}{Christopher Klenk},
  \bibinfo{person}{Karsten K{\"o}nigstein}, \bibinfo{person}{Steffen~P Walz},
  {and} \bibinfo{person}{Arno Schmidt-Trucks{\"a}ss}.}
  \bibinfo{year}{2019}\natexlab{}.
\newblock \showarticletitle{Effectiveness of a behavior change technique--based
  smartphone game to improve intrinsic motivation and physical activity
  adherence in patients with type 2 diabetes: randomized controlled trial}.
\newblock \bibinfo{journal}{\emph{JMIR serious games}} \bibinfo{volume}{7},
  \bibinfo{number}{1} (\bibinfo{year}{2019}), \bibinfo{pages}{e11444}.
\newblock


\bibitem[Kim and Glanz(2013)]%
        {kim2013text}
\bibfield{author}{\bibinfo{person}{Bang~Hyun Kim} {and} \bibinfo{person}{Karen
  Glanz}.} \bibinfo{year}{2013}\natexlab{}.
\newblock \showarticletitle{Text messaging to motivate walking in older African
  Americans: a randomized controlled trial}.
\newblock \bibinfo{journal}{\emph{American Journal of Preventive Medicine}}
  \bibinfo{volume}{44}, \bibinfo{number}{1} (\bibinfo{year}{2013}),
  \bibinfo{pages}{71--75}.
\newblock


\bibitem[Kim et~al\mbox{.}(2018)]%
        {kim2018stretcharms}
\bibfield{author}{\bibinfo{person}{SangBin Kim}, \bibinfo{person}{SinJae Lee},
  {and} \bibinfo{person}{JungHyun Han}.} \bibinfo{year}{2018}\natexlab{}.
\newblock \showarticletitle{StretchArms: Promoting stretching exercise with a
  smartwatch}.
\newblock \bibinfo{journal}{\emph{International Journal of Human--Computer
  Interaction}} \bibinfo{volume}{34}, \bibinfo{number}{3}
  (\bibinfo{year}{2018}), \bibinfo{pages}{218--225}.
\newblock


\bibitem[Kitchenham and Charters(2007)]%
        {kitchenham2007guidelines}
\bibfield{author}{\bibinfo{person}{Barbara Kitchenham} {and}
  \bibinfo{person}{Stuart Charters}.} \bibinfo{year}{2007}\natexlab{}.
\newblock \bibinfo{booktitle}{\emph{Guidelines for performing systematic
  literature reviews in software engineering}}.
\newblock \bibinfo{type}{{T}echnical {R}eport}. \bibinfo{institution}{Keele
  University and Durham University}.
\newblock


\bibitem[Klausen et~al\mbox{.}(2016)]%
        {klausen2016effects}
\bibfield{author}{\bibinfo{person}{Susanne~Hwiid Klausen},
  \bibinfo{person}{Lars~L Andersen}, \bibinfo{person}{Lars S{\o}ndergaard},
  \bibinfo{person}{Janus~Christian Jakobsen}, \bibinfo{person}{Vibeke
  Zoffmann}, \bibinfo{person}{Kasper Dideriksen}, \bibinfo{person}{Anne Kruse},
  \bibinfo{person}{Ulla~Ramer Mikkelsen}, {and} \bibinfo{person}{J{\o}rn
  Wetterslev}.} \bibinfo{year}{2016}\natexlab{}.
\newblock \showarticletitle{Effects of eHealth physical activity encouragement
  in adolescents with complex congenital heart disease: the PReVaiL randomized
  clinical trial}.
\newblock \bibinfo{journal}{\emph{International Journal of Cardiology}}
  \bibinfo{volume}{221} (\bibinfo{year}{2016}), \bibinfo{pages}{1100--1106}.
\newblock


\bibitem[Kooiman et~al\mbox{.}(2018)]%
        {kooiman2018self}
\bibfield{author}{\bibinfo{person}{Thea~JM Kooiman}, \bibinfo{person}{Martijn
  De~Groot}, \bibinfo{person}{Klaas Hoogenberg}, \bibinfo{person}{Wim~P
  Krijnen}, \bibinfo{person}{Cees~P Van Der~Schans}, {and}
  \bibinfo{person}{Adriaan Kooy}.} \bibinfo{year}{2018}\natexlab{}.
\newblock \showarticletitle{Self-tracking of physical activity in people with
  type 2 diabetes: a randomized controlled trial}.
\newblock \bibinfo{journal}{\emph{CIN: Computers, Informatics, Nursing}}
  \bibinfo{volume}{36}, \bibinfo{number}{7} (\bibinfo{year}{2018}),
  \bibinfo{pages}{340--349}.
\newblock


\bibitem[Kramer et~al\mbox{.}(2019)]%
        {kramer2019cluster}
\bibfield{author}{\bibinfo{person}{Jan-Niklas Kramer}, \bibinfo{person}{Peter
  Tinschert}, \bibinfo{person}{Urte Scholz}, \bibinfo{person}{Elgar Fleisch},
  {and} \bibinfo{person}{Tobias Kowatsch}.} \bibinfo{year}{2019}\natexlab{}.
\newblock \showarticletitle{A cluster-randomized trial on small incentives to
  promote physical activity}.
\newblock \bibinfo{journal}{\emph{American journal of preventive medicine}}
  \bibinfo{volume}{56}, \bibinfo{number}{2} (\bibinfo{year}{2019}),
  \bibinfo{pages}{e45--e54}.
\newblock


\bibitem[Lalmas et~al\mbox{.}(2014)]%
        {lalmas2014measuring}
\bibfield{author}{\bibinfo{person}{Mounia Lalmas}, \bibinfo{person}{Heather
  O'Brien}, {and} \bibinfo{person}{Elad Yom-Tov}.}
  \bibinfo{year}{2014}\natexlab{}.
\newblock \showarticletitle{Measuring user engagement}.
\newblock \bibinfo{journal}{\emph{Synthesis Lectures on Information Concepts,
  Retrieval, and Services}} \bibinfo{volume}{6}, \bibinfo{number}{4}
  (\bibinfo{year}{2014}), \bibinfo{pages}{1--132}.
\newblock


\bibitem[Lupton(2014)]%
        {lupton2014self}
\bibfield{author}{\bibinfo{person}{Deborah Lupton}.}
  \bibinfo{year}{2014}\natexlab{}.
\newblock \showarticletitle{Self-tracking cultures: towards a sociology of
  personal informatics}. In \bibinfo{booktitle}{\emph{Proceedings of the 26th
  Australian computer-human interaction conference on designing futures: The
  future of design}}. \bibinfo{pages}{77--86}.
\newblock


\bibitem[Lustria et~al\mbox{.}(2013)]%
        {lustria2013meta}
\bibfield{author}{\bibinfo{person}{Mia Liza~A Lustria}, \bibinfo{person}{Seth~M
  Noar}, \bibinfo{person}{Juliann Cortese}, \bibinfo{person}{Stephanie~K
  Van~Stee}, \bibinfo{person}{Robert~L Glueckauf}, {and} \bibinfo{person}{Junga
  Lee}.} \bibinfo{year}{2013}\natexlab{}.
\newblock \showarticletitle{A meta-analysis of web-delivered tailored health
  behavior change interventions}.
\newblock \bibinfo{journal}{\emph{Journal of health communication}}
  \bibinfo{volume}{18}, \bibinfo{number}{9} (\bibinfo{year}{2013}),
  \bibinfo{pages}{1039--1069}.
\newblock


\bibitem[Matthews et~al\mbox{.}(2016)]%
        {matthews2016persuasive}
\bibfield{author}{\bibinfo{person}{John Matthews}, \bibinfo{person}{Khin~Than
  Win}, \bibinfo{person}{Harri Oinas-Kukkonen}, {and} \bibinfo{person}{Mark
  Freeman}.} \bibinfo{year}{2016}\natexlab{}.
\newblock \showarticletitle{Persuasive technology in mobile applications
  promoting physical activity: a systematic review}.
\newblock \bibinfo{journal}{\emph{Journal of medical systems}}
  \bibinfo{volume}{40}, \bibinfo{number}{3} (\bibinfo{year}{2016}),
  \bibinfo{pages}{72}.
\newblock


\bibitem[Mendoza et~al\mbox{.}(2017)]%
        {mendoza2017fitbit}
\bibfield{author}{\bibinfo{person}{Jason~A Mendoza}, \bibinfo{person}{K~Scott
  Baker}, \bibinfo{person}{Megan~A Moreno}, \bibinfo{person}{Kathryn Whitlock},
  \bibinfo{person}{Mark Abbey-Lambertz}, \bibinfo{person}{Alan Waite},
  \bibinfo{person}{Trina Colburn}, {and} \bibinfo{person}{Eric~J Chow}.}
  \bibinfo{year}{2017}\natexlab{}.
\newblock \showarticletitle{A Fitbit and Facebook mHealth intervention for
  promoting physical activity among adolescent and young adult childhood cancer
  survivors: A pilot study}.
\newblock \bibinfo{journal}{\emph{Pediatric blood \& cancer}}
  \bibinfo{volume}{64}, \bibinfo{number}{12} (\bibinfo{year}{2017}),
  \bibinfo{pages}{e26660}.
\newblock


\bibitem[Michie et~al\mbox{.}(2013)]%
        {michie2013behavior}
\bibfield{author}{\bibinfo{person}{Susan Michie}, \bibinfo{person}{Michelle
  Richardson}, \bibinfo{person}{Marie Johnston}, \bibinfo{person}{Charles
  Abraham}, \bibinfo{person}{Jill Francis}, \bibinfo{person}{Wendy Hardeman},
  \bibinfo{person}{Martin~P Eccles}, \bibinfo{person}{James Cane}, {and}
  \bibinfo{person}{Caroline~E Wood}.} \bibinfo{year}{2013}\natexlab{}.
\newblock \showarticletitle{The behavior change technique taxonomy (v1) of 93
  hierarchically clustered techniques: building an international consensus for
  the reporting of behavior change interventions}.
\newblock \bibinfo{journal}{\emph{Annals of behavioral medicine}}
  \bibinfo{volume}{46}, \bibinfo{number}{1} (\bibinfo{year}{2013}),
  \bibinfo{pages}{81--95}.
\newblock


\bibitem[Monteiro-Guerra et~al\mbox{.}(2019)]%
        {monteiro2019personalization}
\bibfield{author}{\bibinfo{person}{Francisco Monteiro-Guerra},
  \bibinfo{person}{Octavio Rivera-Romero}, \bibinfo{person}{Luis
  Fernandez-Luque}, {and} \bibinfo{person}{Brian Caulfield}.}
  \bibinfo{year}{2019}\natexlab{}.
\newblock \showarticletitle{Personalization in Real-Time Physical Activity
  Coaching using Mobile Applications: A Scoping Review}.
\newblock \bibinfo{journal}{\emph{IEEE Journal of Biomedical and Health
  Informatics}} \bibinfo{volume}{24}, \bibinfo{number}{6}
  (\bibinfo{year}{2019}), \bibinfo{pages}{1738--1751}.
\newblock


\bibitem[M{\"u}ller et~al\mbox{.}(2016)]%
        {muller2016effectiveness}
\bibfield{author}{\bibinfo{person}{Andre~Matthias M{\"u}ller},
  \bibinfo{person}{Stephanie Alley}, \bibinfo{person}{Stephanie Schoeppe},
  {and} \bibinfo{person}{Corneel Vandelanotte}.}
  \bibinfo{year}{2016}\natexlab{}.
\newblock \showarticletitle{The effectiveness of e-\& mHealth interventions to
  promote physical activity and healthy diets in developing countries: a
  systematic review}.
\newblock \bibinfo{journal}{\emph{International Journal of Behavioral Nutrition
  and Physical Activity}} \bibinfo{volume}{13}, \bibinfo{number}{1}
  (\bibinfo{year}{2016}), \bibinfo{pages}{109}.
\newblock


\bibitem[Murnane et~al\mbox{.}(2018)]%
        {Murnane2018}
\bibfield{author}{\bibinfo{person}{Elizabeth~L. Murnane},
  \bibinfo{person}{Tara~G. Walker}, \bibinfo{person}{Beck Tench},
  \bibinfo{person}{Stephen Voida}, {and} \bibinfo{person}{Jaime Snyder}.}
  \bibinfo{year}{2018}\natexlab{}.
\newblock \showarticletitle{{Personal informatics in interpersonal contexts:
  Towards the design of technology that supports the social ecologies of
  long-term mental health management}}.
\newblock \bibinfo{journal}{\emph{Proceedings of the ACM on Human-Computer
  Interaction}} \bibinfo{volume}{2}, \bibinfo{number}{CSCW}
  (\bibinfo{year}{2018}).
\newblock
\showISSN{25730142}
\urldef\tempurl%
\url{https://doi.org/10.1145/3274396}
\showDOI{\tempurl}


\bibitem[Nunes et~al\mbox{.}(2015)]%
        {nunes2015self}
\bibfield{author}{\bibinfo{person}{Francisco Nunes}, \bibinfo{person}{Nervo
  Verdezoto}, \bibinfo{person}{Geraldine Fitzpatrick}, \bibinfo{person}{Morten
  Kyng}, \bibinfo{person}{Erik Gr{\"o}nvall}, {and} \bibinfo{person}{Cristiano
  Storni}.} \bibinfo{year}{2015}\natexlab{}.
\newblock \showarticletitle{Self-care technologies in HCI: Trends, tensions,
  and opportunities}.
\newblock \bibinfo{journal}{\emph{ACM Transactions on Computer-Human
  Interaction (TOCHI)}} \bibinfo{volume}{22}, \bibinfo{number}{6}
  (\bibinfo{year}{2015}), \bibinfo{pages}{1--45}.
\newblock


\bibitem[Oinas-Kukkonen and Harjumaa(2009)]%
        {oinas2009persuasive}
\bibfield{author}{\bibinfo{person}{Harri Oinas-Kukkonen} {and}
  \bibinfo{person}{Marja Harjumaa}.} \bibinfo{year}{2009}\natexlab{}.
\newblock \showarticletitle{Persuasive systems design: Key issues, process
  model, and system features}.
\newblock \bibinfo{journal}{\emph{Communications of the Association for
  Information Systems}} \bibinfo{volume}{24}, \bibinfo{number}{1}
  (\bibinfo{year}{2009}), \bibinfo{pages}{28}.
\newblock


\bibitem[Organization(2019)]%
        {world2019global}
\bibfield{author}{\bibinfo{person}{World~Health Organization}.}
  \bibinfo{year}{2019}\natexlab{}.
\newblock \bibinfo{booktitle}{\emph{Global action plan on physical activity
  2018-2030: more active people for a healthier world}}.
\newblock \bibinfo{publisher}{World Health Organization}.
\newblock


\bibitem[Partners(2014)]%
        {attrition1}
\bibfield{author}{\bibinfo{person}{Endeavour Partners}.}
  \bibinfo{year}{2014}\natexlab{}.
\newblock \bibinfo{booktitle}{\emph{Inside Wearables Part 1: How behavior
  change unlocks long-term engagement}}.
\newblock Endeavour Partners.
\newblock
\urldef\tempurl%
\url{https://medium.com/@endeavourprtnrs/inside-wearable-how-the-science-of-human-behavior-change-offers-the-secret-to-long-term-engagement-a15b3c7d4cf3}
\showURL{%
Retrieved July 28, 2020 from \tempurl}


\bibitem[Prochaska and Velicer(1997)]%
        {prochaska1997transtheoretical}
\bibfield{author}{\bibinfo{person}{James~O Prochaska} {and}
  \bibinfo{person}{Wayne~F Velicer}.} \bibinfo{year}{1997}\natexlab{}.
\newblock \showarticletitle{The transtheoretical model of health behavior
  change}.
\newblock \bibinfo{journal}{\emph{American journal of health promotion}}
  \bibinfo{volume}{12}, \bibinfo{number}{1} (\bibinfo{year}{1997}),
  \bibinfo{pages}{38--48}.
\newblock


\bibitem[Quelly et~al\mbox{.}(2016)]%
        {quelly2016impact}
\bibfield{author}{\bibinfo{person}{Susan~B Quelly}, \bibinfo{person}{Anne~E
  Norris}, {and} \bibinfo{person}{Jessica~L DiPietro}.}
  \bibinfo{year}{2016}\natexlab{}.
\newblock \showarticletitle{Impact of mobile apps to combat obesity in children
  and adolescents: a systematic literature review}.
\newblock \bibinfo{journal}{\emph{Journal for Specialists in Pediatric
  Nursing}} \bibinfo{volume}{21}, \bibinfo{number}{1} (\bibinfo{year}{2016}),
  \bibinfo{pages}{5--17}.
\newblock


\bibitem[Ratten and Ratten(2007)]%
        {ratten2007social}
\bibfield{author}{\bibinfo{person}{Vanessa Ratten} {and}
  \bibinfo{person}{Hamish Ratten}.} \bibinfo{year}{2007}\natexlab{}.
\newblock \showarticletitle{Social cognitive theory in technological
  innovations}.
\newblock \bibinfo{journal}{\emph{European Journal of Innovation Management}}
  \bibinfo{volume}{10}, \bibinfo{number}{1} (\bibinfo{year}{2007}),
  \bibinfo{pages}{90--108}.
\newblock


\bibitem[Roberts et~al\mbox{.}(2017)]%
        {roberts2017digital}
\bibfield{author}{\bibinfo{person}{Anna~L Roberts}, \bibinfo{person}{Abigail
  Fisher}, \bibinfo{person}{Lee Smith}, \bibinfo{person}{Malgorzata Heinrich},
  {and} \bibinfo{person}{Henry~WW Potts}.} \bibinfo{year}{2017}\natexlab{}.
\newblock \showarticletitle{Digital health behaviour change interventions
  targeting physical activity and diet in cancer survivors: a systematic review
  and meta-analysis}.
\newblock \bibinfo{journal}{\emph{Journal of Cancer Survivorship}}
  \bibinfo{volume}{11}, \bibinfo{number}{6} (\bibinfo{year}{2017}),
  \bibinfo{pages}{704--719}.
\newblock


\bibitem[Spanos and Angelis(2016)]%
        {spanos2016impact}
\bibfield{author}{\bibinfo{person}{Georgios Spanos} {and}
  \bibinfo{person}{Lefteris Angelis}.} \bibinfo{year}{2016}\natexlab{}.
\newblock \showarticletitle{The impact of information security events to the
  stock market: A systematic literature review}.
\newblock \bibinfo{journal}{\emph{Computers \& Security}}  \bibinfo{volume}{58}
  (\bibinfo{year}{2016}), \bibinfo{pages}{216--229}.
\newblock


\bibitem[Spiel et~al\mbox{.}(2019)]%
        {spiel2019agency}
\bibfield{author}{\bibinfo{person}{Katta Spiel}, \bibinfo{person}{Christopher
  Frauenberger}, \bibinfo{person}{Os Keyes}, {and} \bibinfo{person}{Geraldine
  Fitzpatrick}.} \bibinfo{year}{2019}\natexlab{}.
\newblock \showarticletitle{Agency of autistic children in technology
  research—A critical literature review}.
\newblock \bibinfo{journal}{\emph{ACM Transactions on Computer-Human
  Interaction (TOCHI)}} \bibinfo{volume}{26}, \bibinfo{number}{6}
  (\bibinfo{year}{2019}), \bibinfo{pages}{1--40}.
\newblock


\bibitem[Strohacker et~al\mbox{.}(2014)]%
        {strohacker2014impact}
\bibfield{author}{\bibinfo{person}{Kelley Strohacker}, \bibinfo{person}{Omar
  Galarraga}, {and} \bibinfo{person}{David~M Williams}.}
  \bibinfo{year}{2014}\natexlab{}.
\newblock \showarticletitle{The impact of incentives on exercise behavior: a
  systematic review of randomized controlled trials}.
\newblock \bibinfo{journal}{\emph{Annals of Behavioral Medicine}}
  \bibinfo{volume}{48}, \bibinfo{number}{1} (\bibinfo{year}{2014}),
  \bibinfo{pages}{92--99}.
\newblock


\bibitem[Thieme et~al\mbox{.}(2020)]%
        {thieme2020machine}
\bibfield{author}{\bibinfo{person}{Anja Thieme}, \bibinfo{person}{Danielle
  Belgrave}, {and} \bibinfo{person}{Gavin Doherty}.}
  \bibinfo{year}{2020}\natexlab{}.
\newblock \showarticletitle{Machine Learning in Mental Health: A Systematic
  Review of the HCI Literature to Support the Development of Effective and
  Implementable ML Systems}.
\newblock \bibinfo{journal}{\emph{ACM Transactions on Computer-Human
  Interaction (TOCHI)}} \bibinfo{volume}{27}, \bibinfo{number}{5}
  (\bibinfo{year}{2020}), \bibinfo{pages}{1--53}.
\newblock


\bibitem[Yfantidou(2020)]%
        {sofia_yfantidou_2020_4063377}
\bibfield{author}{\bibinfo{person}{Sofia Yfantidou}.}
  \bibinfo{year}{2020}\natexlab{}.
\newblock \bibinfo{booktitle}{\emph{syfantid/PAST-SELF-Framework-Data: Paper
  Release}}.
\newblock
\urldef\tempurl%
\url{https://doi.org/10.5281/zenodo.4063377}
\showDOI{\tempurl}


\bibitem[Yfantidou et~al\mbox{.}(2021)]%
        {yfantidou2021self}
\bibfield{author}{\bibinfo{person}{Sofia Yfantidou}, \bibinfo{person}{Pavlos
  Sermpezis}, {and} \bibinfo{person}{Athena Vakali}.}
  \bibinfo{year}{2021}\natexlab{}.
\newblock \showarticletitle{Self-Tracking Technology for mHealth: A Systematic
  Review and the PAST SELF Framework}.
\newblock \bibinfo{journal}{\emph{arXiv preprint arXiv:2104.11483}}
  (\bibinfo{year}{2021}).
\newblock


\bibitem[Zhang and Jemmott~III(2019)]%
        {zhang2019mobile}
\bibfield{author}{\bibinfo{person}{Jingwen Zhang} {and} \bibinfo{person}{John~B
  Jemmott~III}.} \bibinfo{year}{2019}\natexlab{}.
\newblock \showarticletitle{Mobile app-based small-group physical activity
  intervention for young African American women: a pilot randomized controlled
  trial}.
\newblock \bibinfo{journal}{\emph{Prevention Science}} \bibinfo{volume}{20},
  \bibinfo{number}{6} (\bibinfo{year}{2019}), \bibinfo{pages}{863--872}.
\newblock


\end{thebibliography}

% \appendix

\end{document}